\documentclass[comsoc]{IEEEtran}

\usepackage[T1]{fontenc}
\usepackage[cmex10]{amsmath} % Use the [cmex10] option to ensure complicance with IEEE Xplore (see bare_conf.tex)
\usepackage{mathtools} % for underbracket
\usepackage[shortlabels]{enumitem} %customised enumerate
\usepackage{graphicx}
\usepackage[space,noadjust]{cite} % options are chosen such that references appear properly when used in heading in theorem environments
\usepackage{soul,xcolor}
\usepackage[ruled,vlined,linesnumbered]{algorithm2e}
\usepackage{booktabs}
\def\arraystretch{1}
\usepackage{flushend}

\usepackage{etoolbox}
\makeatletter
\patchcmd{\@begintheorem}{\textit}{\textbf}{}{}
\patchcmd{\@begintheorem}{\itshape}{\bfseries}{}{}
\makeatother

\newcommand{\absent}{\mathbb{U}^\text{abs}}
\newtheorem{theorem}{Theorem}
\newtheorem{corollary}{Corollary}
\newtheorem{lemma}{Lemma}

\newtheorem{definition}{Definition}

\newtheorem{remark}{Remark}
\newtheorem{example}{Example}
\newtheorem{condition}{Condition}
\graphicspath{{pictures/}}

\begin{document}

\onecolumn 
\thispagestyle{empty}
\noindent {\Huge IEEE Copyright Notice}

\bigskip

\noindent {\large \copyright~20xx IEEE. Personal use of this material is permitted. Permission from IEEE must be obtained for all other uses, in any current or future media, including reprinting/republishing this material for advertising or promotional purposes, creating new collective works, for resale or redistribution to servers or lists, or reuse of any copyrighted component of this work in other works.}
\newpage
\twocolumn 

%---------- Title ----------
\title{Performance Bounds on Pliable Index Coding Using Absent Receivers}
% Using Absent Receivers to Characterise Pliable Index Coding Rates\\

\author{Lawrence Ong, Badri N.\ Vellambi, Parastoo Sadeghi, J\"{o}rg Kliewer%
\thanks{Lawrence Ong was supported by the Australian Research Council under project FT140100219.}
\thanks{This paper was presented in part at the 2020 International Zurich Seminar on Information and Communication and the 2019 IEEE International Symposium on Information Theory.}
\thanks{This article has been accepted for publication in IEEE Transactions on Information Theory. This is the author's version. The final published version is available in IEEE Xplore. DOI: 10.1109/TIT.2025.3632399}
}

\IEEEoverridecommandlockouts

\maketitle

\begin{abstract}
%\PS{BEGIN NEW}

We characterise bounds on the optimal broadcast rate for a few classes of pliable-index-coding instances. Unlike the majority of currently solved instances, which belong to a special class where all receivers with a certain side-information cardinality are either present or absent, we consider more general instances without this constraint. We devise a novel algorithm that constructs a decoding chain by iteratively adding a message that can be decoded by a receiver whose side information is already in the chain. If the decoding chain cannot proceed due to the absence of a receiver with the required messages, we \textit{skip} a message by adding it to the chain regardless. We prove that a lower bound on the optimal broadcast rate is a function of the number of skipped messages, across all possible decoding choices of the receivers and any realisation of the algorithm for each decoding choice. While this result is not computationally feasible in isolation, it serves as a basis for deriving explicit lower bounds on the broadcast rate for specific classes of pliable-index-coding instances. These lower bounds depend on the number of absent receivers or the pattern of their side-information sets. Specifically, we explicitly characterise the optimal broadcast rate for instances with up to and including four absent receivers with any side-information pattern, as well as instances where the side-information sets are nested in particular ways.

%\PS{END NEW}

% \PS{TWO OLD VERSIONS OF ABSTRACT}
% We characterise the optimal broadcast rate for a few classes of pliable-index-coding problems. This is achieved by devising new lower bounds that utilise the set of \textit{absent} receivers to construct decoding chains with \textit{skipped messages}. This work complements existing works by considering problems that are not complete-$S$, i.e.,  problems considered in this work do not require that all receivers with a certain side-information cardinality to be either present or absent from the problem. We show that for a certain class, the set of receivers is critical in the sense that adding any receiver strictly increases the broadcast rate.

% This paper studies pliable index coding, in which a sender broadcasts information to multiple receivers through a shared broadcast medium, and the receivers each have some message a priori and want any message they do not have. An approach, based on receivers that are absent from the problem, was previously proposed to find lower bounds on the optimal broadcast rate. In this paper, we introduce new techniques to obtained better lower bounds, and derive the optimal broadcast rates for new classes of the problems, including all problems with up to four absent receivers.
\end{abstract}

%%%%
%%%%
\section{Introduction}
%%%%
%%%%
%\PS{BEGIN NEW}

Index coding~\cite{baryossefbirk11} is a canonical open problem in network information theory with a single sender and multiple receivers that are connected via a noiseless broadcast link. Each receiver is characterised by the set of messages that it has, referred to as its side information, and the messages that it wants from the sender. The aim of the index-coding problem is to determine the minimum code rate at which the sender must broadcast to satisfy all receivers, as well as the coding schemes that achieve the optimal rate. While the optimal broadcast rate of index coding remains generally unknown, there has been significant interest in establishing both upper and lower bounds to measure and compare the performance of index codes.
Index coding~\cite{baryossefbirk11,neelytehranizhang13,ongholim16,arbabjolfaeikim18trends}, its secure variant~\cite{dauskachekchee12,ongvellambiyeohklieweryuan2016}, and its connection to network coding~\cite{rouayhebsprintsongeorghiades10,effrosrouayheblangberg15,ongvellambiklieweryeoh21} have received significant research interest.

A variant of index coding, known as \emph{pliable index coding}, was introduced~\cite{brahmafragouli15}, where each receiver does not request a specific subset of messages, but instead is satisfied with \emph{any} subset of $t$ messages it does not already know.
Pliable index coding optimises bandwidth usage for applications where the users are satisfied with any new data that they do not already possess. One such application is push-based content feed for social media, where Facebook, YouTube, Spotify, and Instagram users are happy to receive any relevant new content. Pliable index coding is also useful in distributed machine learning, where the learning algorithm is satisfied with any new data sample from a pool. In software updates, devices can update any patch they do not already have. Similar to the original index-coding problem, the aim is to find the minimum code rate to satisfy pliable demands of all receivers.

Even though index coding and pliable index coding share many similarities, their decoding requirements set them apart in non-trivial ways. As a result, different techniques have been attempted to solve each of them. To date, only a small number of classes of pliable-index-coding problem instances have been solved. In particular, two classes of \textit{symmetrical} instances have been solved~\cite{liutuninetti17, liutuninetti18}. These instances are classified under the complete-$S$ instances, where the receivers are symmetrical in the sense that if a receiver is present in the instance, every receiver with the same cardinality of messages as side information as that of the present receiver is also present.

In more realistic communication scenarios, the set of receivers may not necessarily be symmetrical. For example, if there exists a receiver who already has three messages as side information, it is unlikely that for every combination of three messages, a receiver who has exactly those three messages as side information also exists. In this paper, we will derive bounds on the optimal code rate for asymmetrical cases.

\subsection{Contributions}
%Existing results on exact minimum broadcast rates were established for certain complete-$S$ pliable index coding problems.

% \PS{START NEW}  
In this work, we derive lower bounds on the optimal pliable-index-coding rate and show that they are tight in various settings.
In contrast to previous work, we consider instances that are not complete-$S$, and where each receiver requests exactly $t=1$ message from the sender. Whereas all previous work focused on present receivers in the instance, we focus on how \emph{absent} receivers affect the broadcast rate. We call a receiver $H$ absent if there is no receiver with side information set $H$ in the given instance.

For a pliable-index-coding instance with $m$ messages, the maximum number of receivers that can be present in the system is $2^m -1$, each having a distinct set of messages.\footnote{Note that the receiver who has all $m$ messages cannot be present in the system, as it already knows all messages.} So, looking at absent receivers is just looking at the complement of receivers that are present. Nonetheless, the results in our paper are characterised by the side-information structures of these absent receivers.

We identify a new technique based on absent receivers to derive lower bounds on the optimal broadcast rate for all pliable-index-coding instances that are applicable to both linear and non-linear codes. When combined with matching transmission codes (upper bounds), we establish precisely the optimal broadcast rate for several new classes of pliable-index-coding instances. The lower bounds utilise the concept of \emph{decoding chains with skipped messages}---the fewer the skipped messages, the tighter the lower bound. It turns out that the number and side-information pattern of absent receivers dictate how many messages need to be skipped. Finding the smallest number of skipped messages is a highly non-trivial problem, and its partial characterisation is a main contribution of this paper.  The results of this paper can be summarised as follows.

\begin{itemize}
\item  We first propose Algorithm~\ref{algo:2} that will be used as the basis for deriving our results. The aim of this iterative algorithm, whose properties are proven in Lemma~\ref{lemma:pruning-lower-bound}, is to construct decoding chains and keep track of the skipped messages in the process. Skipping a message may be necessary when the current decoding chain equals (the side information of) an absent receiver. In Theorem~\ref{theorem:chain-lower-bound}, we show that by considering all possible decoding choices and all possible ways of skipping messages, one can find a lower bound on the optimal broadcast rate. Since the number of decoding choices and ways of skipping messages scale exponentially with the number of messages, Algorithm~\ref{algo:2} by itself is not a practical computation method for the lower bound. Nevertheless, we exploit the characteristics of the algorithm to establish new lower bounds as described below. 
 
\item In Theorem~\ref{theorem:simplier-lower-bound}, we establish a lower bound on the optimal broadcast rate that is based on the longest chain of nested absent receivers.\footnote{We say absent receivers $H_1, H_2, \dotsc , H_{L}$ are nested if  $H_1\subsetneq H_2 \subsetneq \cdots \subsetneq H_{L}$.} The insight here is that the number of skipped messages in Algorithm~\ref{algo:2} cannot be larger than the longest chain of absent receivers being sequentially encountered in the process of building any decoding chain. 

\item In Theorem~\ref{theorem:improved-nested-chain}, we utilise a condition under which skipping some messages is ``better'' than others in the sense that the longest chain of absent receivers that can be constructed from that point onward becomes shorter. This then leads to a tighter lower bound compared to Theorem~\ref{theorem:simplier-lower-bound}. The concept of finding good message candidates to skip is further developed in Lemma~\ref{lemma:look-ahead-skip}, which will be used in subsequent theorems. 
%\LO{develop an idea that some skipped messages are ``better'' than others in the sense that the longest chain of absent receivers can be constructed from that point onward is shorter. More specifically, if this can be done for all possible chains of length $L$, then the total skipped messages can be made at most $L-1$ rather than $L$. This then leads to a tighter lower bound. The concept of finding good candidates to skip is further developed in Lemma~\ref{lemma:look-ahead-skip} and in subsequent theorems.} \PS{OLD: utilise a condition under which it is possible to tighten the lower bound obtained in Theorem~\ref{theorem:simplier-lower-bound} by at least one. When the condition in Theorem~\ref{theorem:improved-nested-chain} is met, it becomes possible to skip a certain message in Algorithm~\ref{algo:2} such the effective length of all chains of nested absent receivers of length $L$ is reduced by one.} 

\item In a series of theorems, Theorems~\ref{theorem:incomplete}--\ref{theorem:4-absent}, we establish results for special cases of pliable index coding that involve either special patterns or special numbers of absent receivers. Theorem~\ref{theorem:incomplete} shows that if the union of absent receivers' side information is not equal to the entire message set, then the optimal broadcast rate equals $m-1$, where $m$ is the number of messages. Theorem~\ref{thm:no-absent} strengthens a previous result by proving that the broadcast rate is equal to $m$ if and only if there is no absent receiver in the system. Theorem~\ref{theorem:nesting} specifies the optimal broadcast rate where there is no or exactly one pair of nested absent receivers. Theorems~\ref{theorem:perfectly-nested}--\ref{theorem:truncated-nested} deal with some special patterns of nested absent receivers. Finally, the last results of this paper characterise the optimal broadcast rate for one to four absent receivers.
\end{itemize} 

After reviewing related work in Section~\ref{sec:related}, we present the problem setup in Section~\ref{sec:setup}. Section~\ref{sec:lowerbound} lays the foundations of establishing lower bounds on broadcast rate based on the concept of absent receivers, decoding chains, and skipped messages. It presents  Algorithm~\ref{algo:2}, proves its properties, and establishes a lower bound in Theorem~\ref{theorem:chain-lower-bound}. Based on Theorem~\ref{theorem:chain-lower-bound}, more lower bounds are established in Section~\ref{sec:advanced}, and they are used throughout Section~\ref{sec:application} for studying special classes of absent receivers in pliable index coding. 

\section{Related Works}\label{sec:related}
%%%%
%%%%

To date, the problem of finding optimal broadcast rates and optimal pliable index codes remains open. This holds true even when the problem is restricted to only linear codes, a challenge demonstrated to be NP-hard~\cite{brahmafragouli15}. Nonetheless, bounds on the optimal broadcast rates, as well as algorithms to construct codes, have been developed.

The first upper bound (that is, achievability) on the optimal broadcast rate was derived by Brahma and Fragouli~\cite{brahmafragouli15}. They showed that if each receiver has at least $s_\text{min}$ and at most $s_\text{max}$ messages, then the optimal broadcast rate, $\beta \leq \min \{s_\text{max} + t, m - s_\text{min}\}$, where $m$ is the number of messages and $t$ is the number of new messages that each receiver must get. This upper bound is achieved by either sending (i)~an MDS code of length $m - s_\text{min}$ or (ii)~$s_\text{max} + t$ message uncoded. Since every receiver knows at least $s_\text{min}$ messages, the MDS code allows every receiver to decode all messages. On the other hand, since every receiver knows at most $s_\text{max}$ messages, sending $s_\text{max} + t$ messages uncoded allows it to get at least $t$ new messages. 
% And this bound is tight if the sender knows only the number of messages (as opposed to the exact message sets) that each receiver knows.  For a general setup, they %approximated the behaviour (using the Big-O notation) of the minimum broadcast rate in the limits of the number of messages and receivers.
% approximated the order of dependence of the minimum broadcast rate in the limit as the number of messages and receivers grows.

Song and Fragouli~\cite{songfragouli18} later showed that if all receivers\footnote{By all receivers, we mean for every combination of side information, there exists a receiver who has that combination of messages. An exception is that the receiver who has all messages cannot be present. So, for $m$ messages, the number of all receivers is $2^m-1$.} are present, then the optimal linear broadcast rate is $m$. This is achieved trivially by sending all $m$ messages uncoded. Liu and Tuninetti~\cite{liutuninetti17} strengthened this result to all (including non-linear) pliable index codes. 

Lower bounds have been derived only for a special class of receivers called complete-$S$ instances~\cite{liutuninetti17}, where $S\subseteq \{0,\ldots, m-1\}$ is a parameter.
Given $S\subseteq \{0,\ldots, m-1\}$, the complete-$S$ instance consists of \textit{all} $\binom{m}{i}$ receivers having distinct combination of $i$ messages, for every $i \in S$. 

When $S$ is consecutive in the sense that $S = \{s_\text{min}, \dotsc, s_\text{max}\}$,
%for any $0 \leq s_\text{min} \leq s_\text{max} \leq m-t$, 
a lower bound is given by $\beta \geq \{s_\text{max} + t, m - s_\text{min}\}$~\cite{liutuninetti17}, which coincides with the achievability results by Brahma and Fragouli~\cite{brahmafragouli15}.
When $S$ is complement-consecutive in the sense that $S=\{0,\ldots, m-1\}\setminus\{s_\text{min} ,\ldots, s_\text{max}\} = \{0, \dotsc, S_\text{min}-1\} \cup \{s_\text{max} +1, \dotsc, m-t\}$,
%for any $0 \leq s_\text{min} \leq s_\text{max} \leq m-t$, 
a lower bound is given by $\min \{m, m+t +s_\text{min} - s_\text{max} -2 \}$~\cite{liutuninetti17}. This lower bound also coincides with the achievability using MDS or uncoded schemes~\cite{brahmafragouli15} applied to both partitions of $S$, that is $\min \{s_\text{min} + t - 1, m \} + \min \{m, m - s_\text{max} - 1 \}$, and capping the rate to $m$.

In addition, for a few classes of complete-$S$ instances that are neither consecutive or complement-consecutive, the optimal broadcast rate has also been found to be either achievable by MDS codes or uncoded transmissions.

Besides the MDS codes and uncoded transmissions, which are optimal for several complete-$S$ instances, algorithms to construct codes for the general pliable-index-coding instances have been proposed. They include randomised approaches~\cite{brahmafragouli15} and greedy approaches~\cite{brahmafragouli15,songfragouli18,Eghbal23}. 
These heuristic approaches do not yield precise theoretical achievable rates (upper bounds). Nevertheless, certain approaches yield codes with lengths whose length can be characterised order-wise. Brahma and Fragouli~\cite{brahmafragouli15} used a random-coding method to generate codes of length $\mathcal{O}(\min\{t\log^2{n}, t\log{n} + \log^3{n}\})$ when $m = \mathcal{O}(n^\delta)$, where $n$ represents the number of receivers, $m$ denotes the number of messages, and $\delta$ is a positive constant. Later, Song and Fragouli~\cite{songfragouli18} constructed greedy, deterministic, polynomial-time algorithms that construct linear codes of at most $\frac{2}{\log{1.5}} \log^2{n}$ for $t=1$ and $\mathcal{O}(t \log{n} + \log^2{n})$ for any $t$.

As with index coding~\cite{baryossefbirk11,thapaongjohnson17it}, representing a pliable-index-coding instance with a hypergraph proves to be useful. For hypergraphs with a circular-arc topology, the optimal broadcast rate was shown to be at most 2~\cite{liutuninetti17}. Krishnan et al.~\cite{Krishnan2024} presented a vertex colouring-based approach to pliable index coding. They used the notion of conflict-free colouring of vertices to devise linear pliable index codes. The devised codes marry the conflict-free nature of colouring schemes to successful decoding at the receivers, thereby connecting the broadcast rate to various notions of conflict-free chromatic number. The rate of growth of the codelength of the devised codes is characterised based on the number of messages $m$, the number of demanded messages $t$, and a hypergraph parameter $\Gamma$, which is a measure of the number of receivers whose request set overlaps with that of a receiver. Using specific hypergraph constructions, the devised codes are shown to be order optimal up to a multiplicative factor of $\log t$.

Sowjanya et al.~\cite{SowyanyaITW22} presented another colouring-based linear coding scheme, where the linear combinations of messages are constructed based on independent sets of edges in the hypergraph representation of the pliable-index-coding formulation. Using this approach, the optimal codelength is bounded from above by the maximum degree of vertices in the hypergraph. A lower bound to the optimal codelength is based on the \emph{nesting number} of the hypergraph; the nesting number is the maximal number of nested hyperedges that form a binary tree with a specific nesting structure.

In pliable index coding with $m$ messages and $n$ clients, where $m = \mathcal{O}(n^\delta)$ for some constant $\delta$, and each client possesses \textit{random} side information---specifically, each message is available to each client independently with probability $p$---Song and Fragouli~\cite{songfragouli18} showed that the optimal broadcast rate is almost surely $\Theta(\log{n})$.

%For pliable index coding over random graph B(m,n, p) (m = O(nδ) for some constant δ) with constant p, we can achieve the optimal linear pliable index code length K = (log(n)) almost surely.

A variant of pliable index coding has also been considered in the literature; in this variant, instead of requesting any unknown messages, each receiver specifies its preference over the unknown messages. Heuristic algorithms that consider both the broadcast rate and the overall receiver satisfaction have been proposed~\cite{songfragouli18b,byrneongsadeghivellambi23}.

%%%%
%%%%

\section{Problem Setup}\label{sec:setup}
%%%%
%%%%

%\subsection{Problem Formulation}\label{sec:formulation}

We use the following notation: $\mathbb{Z}^+$ denotes the set of natural numbers, $[a:b] := \{a, a+1, \dotsc, b\}$ for $a,b\in\mathbb{Z}^+$ such that $a \leq b$, and $X_S = (X_s: s \in S)$ for some ordered set $S$.

A pliable-index-coding instance consists of the following:  A sender has $m \in \mathbb{Z}^+$ messages, denoted by $X_{[1 : m]} = (X_1, \dots, X_m)$. Each message $X_i \in \mathbb{F}_q$  is independently and uniformly distributed over a finite field of size~$q$. There are $n \in [1:2^m - 1]$ receivers having distinct subsets of messages, which we refer to as side information. Each receiver is labelled by its side information, i.e., the receiver that has messages $X_{H}$, for some $H \subsetneq [1 : m]$, will be referred to as receiver $H$ or receiver with side information $H$. Let $\mathbb{U}$ be the set of all side information sets of the receivers.

We aim to devise an encoding scheme for the sender and a decoding scheme for each receiver, satisfying pliable recovery of $t=1$ message at each receiver.
Denote a pliable-index-coding instance with $m$ messages and $\mathbb{U}$ receivers by $\mathcal{P}_{m,\mathbb{U}}$. A pliable index code of length $\ell \in \mathbb{Z}^+$ for $\mathcal{P}_{m,\mathbb{U}}$ consists of
\begin{itemize}
\item an encoding function of the sender, $\mathsf{E}: \mathbb{F}_q^m \rightarrow \mathbb{F}_q^\ell$,
\item a decoding choice, $D: \mathbb{U} \rightarrow [1:m]$,
\item for each receiver $H\in\mathbb{U}$,
a decoding function $\mathsf{G}_H: \mathbb{F}_q^\ell \times \mathbb{F}_q^{|H|} \rightarrow \mathbb{F}_q$,
\end{itemize}
such that
\begin{align}
    & D(H) \in [1:m] \setminus H,\\
    & \mathsf G_H(\mathsf{E}(X_{[1:m]}),X_{H}) = X_{D(H)}.
\end{align}
for all $H \in \mathbb{U}$ and for all $X_{[1:m]} \in \mathbb{F}_q^m$.
Here, $D(H)$ is the index of the message decoded by receiver $H$, and $\ell$ is the code length. For the rest of the paper, when the context is clear, we may refer to message~$X_i$ simply as message~$i$ instead of the message indexed by $i$.

The aim is to find the optimal broadcast rate for a particular message size $q$, denoted by $$\beta_q := \min_{\mathsf{E}, D, \{\mathsf{G}\}} \ell,$$ and the optimal broadcast rate over all $q$, denoted by $$\beta := \inf_q \beta_q.$$

Without loss of generality, the side-information sets of the receivers are distinct. This is because all receivers having the same side information can be satisfied by the same code---with the same $\mathsf{E}, D, \{\mathsf{G}\}$---if and only if (iff) any one of them can be satisfied. This is why each receiver can be identified by its side information set. Also, no receiver has side information~$H = [1:m]$ because this receiver cannot be satisfied, that is, $D(H)$ cannot be defined. Thus $\mathbb{U} \subseteq 2^{[1:m]} \setminus \{[1:m]\}$. Lastly, %a receiver is said to be \textit{absent} if it is not present in the problem.
any receiver that is not present, that is, receiver~$H$ where $H \neq [1:m]$ and $H \notin \mathbb{U}$, is said to be \textit{absent}. Denote the set of  absent receivers by $\absent := 2^{[1:m]} \setminus (\{[1:m]\} \cup \mathbb{U})$.

\begin{example}
Let $m= 3$, and  $\mathbb U=\{\emptyset, \{1\}, \{2\}, \{1,2\}, \{2,3\}\}$. Then, the receivers $\absent = \{\{3\}, \{1,3\} \}$ are absent.
\end{example}

\begin{remark}
  All results in this paper will be derived for $\beta_q$ for all $q \in \mathbb{Z}^+$. Consequently, the results independent of $q$ are also valid for $\beta$.
\end{remark}

A trivial lower bound on the optimal broadcast rate can be obtained by simply removing receivers.
Since a pliable index code for a receiver collection $\mathbb{U}$ also works for a receiver collection that is a subset of $\mathbb{U}$, the following holds:

\begin{lemma} \label{lemma:monotonicity}
Let $\mathbb U^{-}\subseteq \mathbb U$. Then, $\beta_q (\mathcal{P}_{m,\mathbb{U}^-}) \leq \beta_q(\mathcal{P}_{m,\mathbb{U}})$.
%   Consider two pliable-index-coding problem $\mathcal{P}_{m,\mathbb{U}}$ and $\mathcal{P}_{m,\mathbb{U}^-}$, where $\mathbb{U}^- \subseteq \mathbb{U}$, meaning that some receivers may be missing in the second problem. We have that
  % \begin{equation}
  %   \beta_q (\mathcal{P}_{m,\mathbb{U}^-}) \leq \beta_q(\mathcal{P}_{m,\mathbb{U}}). \label{eq:monotonic}
  % \end{equation}
 \end{lemma}

% \begin{IEEEproof}
%   Any pliable index code for $\mathcal{P}_{m,\mathbb{U}}$ is also a pliable index code for $\mathcal{P}_{m,\mathcbb{U}^-}$.
% \end{IEEEproof}

\section{Lower Bounds Based on Absent Receivers}\label{sec:lowerbound}
%%%%
%%%%

In this section, we will devise an algorithm that outputs lower bounds for any instance of pliable index coding. The algorithm is based on receivers that are absent in the instance, that is, $\absent$. As the algorithm builds on an existing result on index coding, we start by revisiting these relevant results.

\subsection{Linking Pliable Index Coding to Index Coding}

In this section, we derive lower bounds on the optimal broadcast rate for pliable-index-coding instances with lower bounds on the optimal broadcast rates for index-coding instances with the same sets of messages and receivers. Recall that index coding differs from its pliable version by requiring each receiver to decode a specific message. This information is captured by the decoding choice~$D$ defined in Section~\ref{sec:setup}:
\begin{equation}
  D: \mathbb{U} \rightarrow [1:m], \text{ such that } D(H) \in [1:m] \setminus H.
\end{equation}
Here, $D(H)$ is the message that receiver~$H$ decodes.

%the set of receivers by $\mathbb{U} := \{H \subsetneq [1:m]: \text{receiver $H$ is present.}\}$. Define \textit{decoding choice} vector $D= (d_H: H \in \mathbb{U})$, where $d_H \in [1:m] \setminus H$ is the index of the message decoded by receiver $H$. The ordering of $d_H$ in $D$ can be fixed by, say, increasing $i(H) = \sum_{i=1}^m \bm{1}_H(i)2^i$, where $\bm{1}_H(i)$ is the indicator function.

Denote by $\mathcal{P}_{m,\mathbb{U},D}$ a pliable-index-coding instance $\mathcal{P}_{m,\mathbb{U}}$ with a fixed decoding choice~$D$. 
$\mathcal{P}_{m,\mathbb{U},D}$ is in fact an index-coding instance~\cite{neelytehranizhang13,ongholim16,arbabjolfaeikim18trends}, %An index-coding problem is defined by a set of messages $X_{[1:m]}$ and a set of receivers $R$, each receiver~$i \in R$ has  messages $S_i \subsetneq [1:m]$ and wants to decode messages  $W_i \subseteq [1:m] \setminus S_i$.
%By mapping $\mathcal{P}_{m,\mathbb{U},D}$ to an index-coding problem $\mathcal{I}_{m,\mathbb{U},D}$ 
with a message set~$X_{[1:m]}$ and a receiver set~$\mathbb{U}$, where each receiver~$H \in \mathbb{U}$ has $X_H$ and wants $X_{D(H)}$.
% , we have the following:

So, any encoding function $\mathsf{E}$ and decoding functions with the restriction that $\mathsf{D}_H(\mathsf{E}(X_{[1:m]}),X_H) = X_{D(H)}$ for all $H \in \mathbb{U}$ for $\mathcal{P}_{m,\mathbb{U}}$ is a code for $\mathcal{P}_{m,\mathbb{U},D}$, and vice versa.
With an abuse of notation, let the optimal broadcast rate for $\mathcal{P}_{m,\mathbb{U},D}$ be $\beta_q(\mathcal{P}_{m,\mathbb{U},D})$.
We establish the following:
\begin{lemma}\label{lemma:equivalence}
  $\beta_q (\mathcal{P}_{m,\mathbb{U}}) = \min_D \beta_q(\mathcal{P}_{m,\mathbb{U},D}).$
\end{lemma}

\begin{IEEEproof}
 Clearly,
%\begin{equation}
  $\beta_q (\mathcal{P}_{m,\mathbb{U}}) \leq \beta_q(\mathcal{P}_{m,\mathbb{U},D})$
%\end{equation}
for all $D$ because any code for $\mathcal{P}_{m,\mathbb{U},D}$ is a code for $\mathcal{P}_{m,\mathbb{U}}$. Since the inequality must be tight for at least one $D$, we have Lemma~\ref{lemma:equivalence}.
%\begin{equation}
%$\beta_q (\mathcal{P}_{m,\mathbb{U}}) = \min_D \beta_q(\mathcal{P}_{m,\mathbb{U},D})$.
% ,
% %\end{equation}
% where the term on the right-hand side can be written as
% %\begin{equation}
%   $\beta_q(\mathcal{P}_{m,\mathbb{U},D}) = \beta_q(\mathcal{I}_{m,\mathbb{U},D})$,
% %\end{equation}
% because any pliable index code for $\mathcal{P}_{m,\mathbb{U},D}$ is an index code for $\mathcal{I}_{m,\mathbb{U},D}$, and vice versa.
\end{IEEEproof}

From Lemma~\ref{lemma:equivalence}, $\beta_q (\mathcal{P}_{m,\mathbb{U}})$ can be obtained by evaluating the optimal broadcast rates $\beta_q(\mathcal{P}_{m,\mathbb{U},D})$ of index-coding instances~$\mathcal{P}_{m,\mathbb{U},D}$ for all $D$. However, the optimal broadcast rate for index coding is not known in general, and the search space over all possible $D$ grows as $\prod_{H \in \mathbb{U}} (m - |H|)$. %\PS{check this.}\LO{Updated this.}

\subsection{A Lower Bound for Pliable Index Coding Based on Acyclic Subgraphs}

Ignoring computational complexity for now, we will utilise Lemma~\ref{lemma:equivalence} to formulate a lower bound for $\beta_q (\mathcal{P}_{m,\mathbb{U}})$ using results from index coding. More specifically,
\begin{equation} \label{eq:linking-PIC-to-IC}
  \beta_q (\mathcal{P}_{m,\mathbb{U}}) = \min_D \beta_q(\mathcal{P}_{m,\mathbb{U},D}) \geq \min_D \phi_q(\mathcal{P}_{m,\mathbb{U},D}),
\end{equation}
where $\phi_q(\mathcal{P}_{m,\mathbb{U},D})$ is any lower bound on $\beta_q(\mathcal{P}_{m,\mathbb{U},D})$ for the index-coding instance $\mathcal{P}_{m,\mathbb{U},D}$.

We now state a lower bound for index coding~\cite{neelytehranizhang13}, expressed through a directed-bipartite-graph representation of an index-coding instance, defined as follows:

\begin{definition}
  An index-coding instance $\mathcal{P}_{m,\mathbb{U},D}$ can be described by the following bipartite graph:
  \begin{itemize}
      \item Two independent sets:
      \begin{itemize}
                \item A receiver set $\mathbb{U}$.
          \item A message set $[1:m]$.
      \end{itemize}
      \item Directed edges between the independent sets: For every receiver~$H \in \mathbb{U}$, there is
      \begin{itemize}
          \item An edge from receiver~$H$ to every message~$\{i \in [1:m]: i \in H\}$, denoting the messages that the receiver has. 
          \item An edge from message~$D(H) \in [1:m] \setminus H$ to receiver~$H$, denoting the message that the receiver must decode.
          \end{itemize}
  \end{itemize}
\end{definition}

%Any index-coding problem can be specified by a bipartite graph with these two disjoint, independent sets: the message node set and the receiver node set. A directed edge from receiver node~$r$ to message node~$m$ exists iff receiver~$r$ has $X_m$ as side information; a directed edge from message node~$m$ to receiver node~$r$ exists iff receiver~$r$ wants $X_m$. 

The desired lower bound on the optimal broadcast rate of an index-coding instance $\mathcal{P}_{m,\mathbb{U},D}$ will be stated in terms of a bipartite graph~$G'$ as a result of performing a series of \textit{pruning operations} on the bipartite graph~$G$ that describes $\mathcal{P}_{m,\mathbb{U},D}$. The pruning operations comprise one or more of the following steps, repeated as many times as desired and in any order:
\begin{enumerate}
    \item Remove a message node and all its incoming and outgoing edges.
    \item Remove a receiver node and all its incoming and outgoing edges.
    \item Remove a message-to-receiver edge. 
\end{enumerate}

Let $G'$ be the resultant bipartite graph after a series of pruning operations and $m(G')$ denote the \textit{number of message nodes} each with at least one outgoing edge. If $G'$ is acyclic (in the directed sense), then we have the following lower bound:
\begin{lemma}{\cite[Lemma~1]{neelytehranizhang13}} \label{lemma:index-coding-lower-bound} Consider an index-coding instance~$\mathcal{P}_{m,\mathbb{U},D}$ and its bipartite-graph representation~$G$.
  After a series of pruning operations, if the resultant graph~$G'$ is acyclic, then
  %\begin{equation}
  $\beta_q(\mathcal{P}_{m,\mathbb{U},D}) \geq m(G')$.
%\end{equation}
\end{lemma}

Note that for any $G$, we can always find at least one series of pruning operations to obtain an acyclic $G'$. A trivial case is to remove all but one receiver. Also, note that the above lower bound generalises the maximum-acyclic-induced-subgraph (MAIS) lower bound~\cite{baryossefbirk11}.

Combining Lemmas~\ref{lemma:equivalence} and \ref{lemma:index-coding-lower-bound},  we can lower bound the optimal rate for a pliable-index-coding instance in terms of a lower bound for its associated index-coding instances:
\begin{lemma}\label{lemma:acyclic}
  Consider a pliable-index-coding instance $\mathcal{P}_{m,\mathbb{U}}$ and a set of bipartite graphs $\{G_D\}$ formed by all possible decoding choices $D$. Perform pruning operations on each $G_D$ to obtain an acyclic $G_D'$.  Then,
  \begin{equation}
    \beta_q (\mathcal{P}_{m,\mathbb{U}}) \geq \min_D m(G_D').
  \end{equation}
\end{lemma}

\begin{algorithm}[ht]
\SetKwInOut{Input}{input}
\SetKwInOut{Output}{output}

\Input{$\mathcal{P}_{m,\mathbb{U},D}$} 
\Output{A \textit{decoding chain} $C$ (a totally ordered set with a total order $\preceq$) and  a set of \textit{skipped messages} $S$}
$C \leftarrow \emptyset$; \texttt{\scriptsize\color{blue} (initialise $C$)}\\
$S \leftarrow \emptyset$; \texttt{\scriptsize\color{blue} (initialise $S$)}\\
\While{$C \neq [1:m]$}{
  \If(\texttt{\scriptsize\color{blue} (receiver~$C$ is present)}){$C \in \mathbb{U}$}{
    $a \leftarrow D(C)$;
    }

  \Else(\texttt{\scriptsize\color{blue} (receiver~$C$ is absent)}){
    
  \If{$\exists B \in \mathbb{U}$ such that $B \subsetneq C$ and $D(B) \notin C$}{
  Arbitrarily choose one of the following options:\\
    \SetAlgoVlined
    \SetKwProg{Fn}{Option}{:}{}
    \Fn(\texttt{\scriptsize\color{blue} (skip a message)}){\textbf{\upshape 1}}{
              Arbitrarily choose any $a \in [1:m] \setminus C$; \\
    $S \leftarrow S \cup \{a\}$; \texttt{\scriptsize\color{blue} (skip $a$)}\\

  }
  \Fn(\texttt{\scriptsize\color{blue} (avoid skipping)}){\textbf{\upshape 2}}{
Arbitrarily choose any present receiver~$B$ that satisfies the condition in line 7;\\
$a \leftarrow D(B)$;
}
    }

  \Else
  {
    Execute \textbf{Option~1} above.
  }
  }
          $C \leftarrow C \cup \{a\}$; \texttt{\scriptsize\color{blue} (expand the decoding chain $C$)}\\
       Define $i \preceq a,$ for all $i \in C \setminus \{a\}$;  \texttt{\scriptsize\color{blue} (define order in $C$)}\\ 
  }
\caption{An algorithm to construct a decoding chain with skipped messages}
\label{algo:2}
\end{algorithm}

\subsection{An Algorithm to Find Acyclic Subgraphs Using Decoding Chains with Skipped Messages}

\subsubsection{The basic idea of decoding chains}

To use Lemma~\ref{lemma:acyclic}, one needs to consider all $D$ and perform pruning operations on each $G_D$ to get an acyclic graph $G_D'$ and the associated $m(G_D')$. %Note that calculating $m(G_D')$ for only a subset of $D$ does not necessarily give a lower bound to the pliable-index-coding problem.

We will construct an algorithm, Algorithm~\ref{algo:2}, to find $m(G_D')$, which uses the concept of \textit{decoding chain}. The concept of decoding chain was used to prove the MAIS lower bound~\cite{baryossefbirk11} and its extension~\cite[Lemma~1]{neelytehranizhang13} for index coding, and lower bounds for complete-$S$ pliable-index-coding instances~\cite{liutuninetti17,liutuninetti18}.

Consider the following example of a decoding chain: Suppose that receivers $H_1 = \emptyset$, $H_2 = \{a\}$, and $H_3 = \{a,b\}$ are present, and $D$ is chosen such that $D(H_1) = a$, $D(H_2) = b$, and $D(H_3) = c$. After decoding $a$, receiver~$H_1$ (who has as much side information as receiver~$H_2$ has) must be able to decode message $b$, which is the message $H_2$ can decode. Using the same argument, $H_1$ can also decode message~$c$. So, the length of the codeword must be bounded from below as $\ell \geq 3$, since receiver $H_1$, who has no side information, can decode three messages from the transmitted codeword~$\mathsf{E}(X_{[1:m]})$.

\subsubsection{The basic idea of skipping messages}

However, the above-mentioned decoding-chain argument may not proceed if any of the required receivers is not present. To handle this scenario, we propose in Algorithm~\ref{algo:2} a new approach to constructing decoding chains.

Consider the above-mentioned example again but with the following modification: Only $H_1 = \emptyset$ and $H_3 = \{a,b\}$ are present, and $D$ is chosen such that $D(H_1) = a$ and $D(H_3) = c$. The previous argument we used to establish $\ell \geq 3$ fails, as having message~$a$ now does not guarantee the decoding of $b$, and so $H_1$ cannot \textit{mimick} $H_3$ to decode $c$.

To solve this, we \textit{provide} receiver $H_1$ with an additional message~$b$, and this allows the decoding-chain argument to continue for $H_1$ to decode $c$. So, with the codeword~$\mathsf{E}(X_{[1:m]})$ and the additional message~$b$, $H_1$ can decode $\{a,b,c\}$. This gives $H(\mathsf{E}(X_{[1:m]})) + H(X_b) \geq H(\mathsf{E}(X_{[1:m]}), X_b) \geq H(X_a, X_b, X_c)$, resulting in the lower bound $\ell \geq 2$.

In Algorithm~\ref{algo:2}, the act of providing such additional messages is implemented as \textit{skipped} messages, and each skipped message reduces the lower bound by one.

\subsubsection{The algorithm}

Algorithm~\ref{algo:2} runs in iterations. It starts with an empty ordered message subset~$C$. During each iteration, one message is added to $C$. This continues until $C = [1:m]$ at which point the algorithm ends.

The procedure for building $C$ is as follows: We check if knowing the message subset $C$ allows any decoder to decode another message not in $C$. This is possible if (i) receiver~$C$ is present or (ii) receiver~$B \subsetneq C$ is present and $D(B) \notin C$. In either case, $D(C)$ or $D(B)$ can be added to $C$.

In the case that knowing $C$ does not allow a receiver to decode another message (by mimicking another receiver), the algorithm ``skips'' a message~$a \notin C$ by adding $a$ to both the skipped message set~$S$ and the decoding chain~$C$. Note that the choice of $a$ is arbitrary and is not dictated by the decoding choice~$D$.

The number of skipped messages $|S|$ keeps track of the number of ``break points'' in the decoding chain.
We will see later in Lemma~\ref{lemma:pruning-lower-bound} that the optimal pliable index code rate is bounded from below by $m - |S|$. So, each skipped message can be seen as a ``penalty'' to the lower bound.

We say that the algorithm ``\textit{hits}'' a receiver $H$ whenever the ordered message subset~$C$ is updated as $C \leftarrow H$. If receiver~$H$ is absent, we say that it hits an absent receiver~$H$. Note that receiver $[1:m]$ cannot exist, so when the algorithm ends, $[1:m]$ is not considered an absent receiver being hit.

When the algorithm hits an absent receiver~$C$ during the iteration, if receiver~$B \subsetneq C$ is present and $D(B) \notin C$ (that is, the condition on line~7 of the algorithm is satisfied), the algorithm can either skip a message or add $D(B)$ to $C$ without skipping any message. We call the two options Option~1 and Option~2, respectively. These options will be discussed in Section~\ref{sec:algo-lower-bound}.%Both options yield a valid lower bound of $m - |S|$.

% \LO{even if the condition on line~7 of the algorithm is satisfied (that is, it is possible to add a message to $C$ without skipping any message),} the algorithm can arbitrarily decide to skip a message by adding an arbitrary $a \notin C$ to $C$. Doing this will still yield a valid lower bound $m - |S|$.
% \LO{When an absent receiver is hit, we call the choice of skipping a message Option~1 and the choice of not skipping a message Option~2. Note that Option~1 is always possible upon hitting an absent receiver, but Option~2 further requires the condition on line~7 to be met.}
%We call this Option~1 in the algorithm. Some of the results are derived based on Option~1.

\subsubsection{Acyclic subgraph via Algorithm~\ref{algo:2}}

Let $(C,S)$ denote the output of one realisation of Algorithm~\ref{algo:2}, and $\mathbb{C}$, the set of all $(C,S)$ pairs constructed by all possible realisations of the algorithm. We have the following:

\begin{lemma} \label{lemma:pruning-lower-bound}
Consider a given $\mathcal{P}_{m,\mathbb{U},D}$ (or equivalently, the bipartite graph representation $G_D$). For each $(C,S) \in \mathbb{C}$ obtained by Algorithm~\ref{algo:2}, there exists a series of pruning operations on $G_D$ yielding an acyclic $G_D'$ with $m(G_D') = m - |S|$.
\end{lemma}

\begin{IEEEproof}
Note that for each $c \in C \setminus S$, message~$c$ is to be decoded by some receiver~$H$ (see lines 5, 14, and 17). Denote this receiver by $D^{-1}(c)$. While there can be multiple receivers whose decoding choice is message~$c$, here, $D^{-1}(c)$ is used to denote the specific one chosen for a particular realisation of the algorithm, and is hence unique.

We construct the required pruning operations on $G_D$ as follows:
%%% worked until here %%%
\begin{itemize}
    \item Remove all receiver nodes (and all their incoming and outgoing edges), except $\{D^{-1}(c) \in \mathbb{U}: c \in C \setminus S\}$. As the algorithm ends with $|C| = m$, we have $|C \setminus S| = m - |S|$. All receivers in $\{D^{-1}(c) \in \mathbb{U}: c \in C \setminus S\}$ must be distinct as each receiver only decodes one message. So, the number of receivers that remain is $m - |S|$. 
    \item Remove all message nodes (and all their incoming and outgoing edges), except those in  $C \setminus S$. So,  $m - |S|$ messages remain. 
\end{itemize}

The resultant graph $G_D'$ has these remaining edges for each remaining receiver~$H \in \{D^{-1}(c) \in \mathbb{U}: c \in C \setminus S\}$:
\begin{itemize}
    \item Outgoing edges from receiver node~$H$ to each message node~$i \in H \setminus S$. This is because by definition, receiver $H$ knows messages in $H$, but message nodes in $S$ have been removed from the graph.
    \item An incoming edge from message node~$D(H)$ to receiver node~$H$.
\end{itemize}
By construction (see line 18), for each remaining receiver node~$H$, $i \preceq D(H)$ for all $i \in H$. %, where  $c_i \leq_C c_j$ iff $i \leq j$.

% Remove from $G_D$ all present receivers not being hit in the algorithm, and their connected edges.
% Let the elements of $C$ in the order of construction of $C$ be $c_1,c_2,\ldots, \underline{c_i}, \ldots, c_{|C|}$, that is, $c_i \leq_C c_j$ iff $i \leq j$, where underlined elements are present in $S$ as well. By construction, if $c_i$ is underlined, then receiver~$\{c_1, \dotsc, c_{i-1}\}$ is absent. So, for each $c_i$ in $C$ that is not underlined, receiver~$\{c_1, \dotsc, c_{i-1}\}$ is present and has been hit in the algorithm, and therefore remains. This includes receiver~$\emptyset$ if $c_1$ is not underlined. So,  $|C \setminus S|$ receivers remain.
% Next, remove all messages in $S$ (and their associated edges) so that only messages in $C \setminus S$ remain.

% After these pruning operations, the graph $G_D'$ consists of the following edges for each remaining receiver node~$H$: \textsf{(a)}~outgoing edges from $H$ to all message nodes $i \in H \setminus S$, \textsf{(b)}~incoming edge from message node $D(H)$ to $H$. Also, by construction, for each remaining receiver node~$H$, $i \leq_C D(H)$ for all $i \in H$.

For $a \preceq b$, we say that $b$ is \textit{larger} than $a$, and $a$ is \textit{smaller} than $b$.
  In $G_D'$, all edges flow from message nodes that are larger to message nodes that are smaller, through receiver nodes. Hence, $G_D'$ is acyclic. Also, since each message node that remains is requested by a receiver that remains, it has an outgoing edge. So, $G_D'$ contains $m - |S|$ message nodes.
\end{IEEEproof}

\subsubsection{A realisation of the algorithm}
\begin{figure}
    \centering
\includegraphics[width=0.8\linewidth]{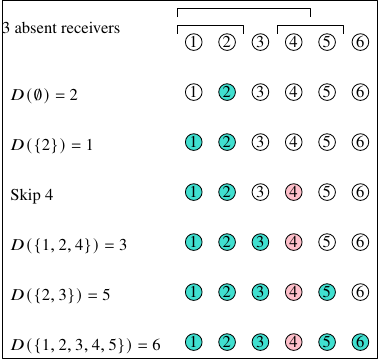}
\caption{A realisation of Algorithm~\ref{algo:2} in Example~\ref{ex:realisation-of-algorithm}, where each row indicates an iteration, the coloured nodes are $C$ after the iteration, and the red nodes are $S$}
\label{fig:algo-2-example}
\end{figure}

\begin{example} \label{ex:realisation-of-algorithm}
    Fig.~\ref{fig:algo-2-example} shows a realisation of the algorithm for the pliable-index-coding instance with $m=6$ messages and three absent receivers $\absent = \{ \{1,2\}, \{1,2,3,4\}, \{4,5\} \}$. The iterations of a possible realisation of the algorithm can be as follows:
\begin{enumerate}[i.]
    \item The algorithm starts with $C = S = \emptyset$. Suppose that $D(\emptyset) = 2$. So, $C$ and $S$ are updated as $C = \{2\}$ and $S = \emptyset$.
    \item Suppose that $D(\{2\}) = 1$. So, $C$ and $S$ are updated as $C = \{2,1\}$ and $S = \emptyset$.
    \item Since receiver $C=\{2,1\}$ is absent, the algorithm can skip a message (using Option 1).  Suppose that it skips message~4. Then, $C$ and $S$ are updated as $C = \{2,1,4\}$ and $S = \{4\}$.
    \item Suppose that $D(\{1,2,4\}) = 3$. So, $C$ and $S$ are updated as $C = \{2,1,4,3\}$ and $S = \{4\}$.
    \item Now, receiver $C = \{2,1,4,3\}$ is absent. But note that receiver $D(\{2,3\})$ is present and suppose that $D(\{2,3\}) = 5 \notin C$. The algorithm can avoid skipping a message (using Option 2), and update $C$ and $S$ to be $C = \{2,1,4,3,5\}$ and $S = \{4\}$.
    \item Lastly, $D(\{1,2,3,4,5\})=6$, and the algorithm terminates with $C = \{2,1,4,3,5,6\}$ and $S = \{4\}$.
\end{enumerate}
\end{example}

%\PS{At this point it will be good to give an example of algorithm and how the acyclic graph is formed.}

\subsection{A Lower Bound for Pliable Index Coding via Algorithm~\ref{algo:2}} \label{sec:algo-lower-bound}

Combining Lemmas~\ref{lemma:acyclic} and \ref{lemma:pruning-lower-bound}, we get the following lower bound obtained by Algorithm~\ref{algo:2}. The lower bound is obtained by taking the maximum number of skipped messages from running Algorithm~\ref{algo:2} over all decoding choices~$D$.
\begin{theorem} \label{theorem:chain-lower-bound}
  For any pliable-index-coding instance $\mathcal{P}_{m,\mathbb{U}}$,
  \begin{equation}
    \beta_q (\mathcal{P}_{m,\mathbb{U}}) \geq  m - \max_D  |S|, \label{eq:chain-lower-bound-not-opt}
  \end{equation}
  where $S$ is the set of skipped messages resulting from a realisation of Algorithm~\ref{algo:2} for a fixed $D$.
\end{theorem}

The lower bound is obtained by maximising $|S|$ over all decoding choices $D$. By optimising the choice of skipped messages for each $D$ such that the minimum number of messages is skipped, we obtain the following lower bound:
\begin{equation}
    \beta_q (\mathcal{P}_{m,\mathbb{U}}) \geq  m - \max_D \min_{(C,S) \in \mathbb{C}}  |S| = m - L^*, \label{eq:chain-lower-bound-new-2}
  \end{equation}
  where we define
  \begin{equation}
  L^* := \max_D \min_{(C,S) \in \mathbb{C}} |S|. \label{eq:chain-lower-bound-b}
\end{equation}

% By optimising $(C,S) \in \mathbb{C}$ of the algorithm for each $D$, the lower bound in Lemmma~\ref{theorem:chain-lower-bound} is optimised as
%   \begin{equation}
%     \beta_q (\mathcal{P}_{m,\mathbb{U}}) \geq  m - \max_D  \min_{(C,S) \in \mathbb{C}} |S|. \label{eq:chain-lower-bound}
%   \end{equation}

The tightest lower bound in Theorem~\ref{theorem:chain-lower-bound} is attained by minimising the set of skipped messages for each decoding choice, c.f. \eqref{eq:chain-lower-bound-not-opt}. This suggests that not skipping a message is preferred over skipping one. 
Therefore, whenever the algorithm hits an absent receiver~$C$, the condition of whether there exists some $B \subsetneq C$ for which $D(B) \notin C$ could be checked (see line~7 in the algorithm). If the condition is true, the algorithm could choose Option~2 to add $D(B)$ to $C$ without skipping any message, rather than executing Option~1 to skip a message. 

However, the lower bound \eqref{eq:chain-lower-bound-not-opt} depends on the output of realisations of Algorithm~\ref{algo:2} over all possible decoding choices, and this can complicate analysis. As such, Option~1 has been used to prove certain results in this paper, as it obviates the check for the existence of some $B \subsetneq C$ for which $D(B) \notin C$ whenever an absent receiver is hit---such a check can potentially lead to different cases to handle for different decoding choices $D$.

In summary, Option~2 is preferred in general, but Option~1 can be used if we want to avoid checking for the existence of some $B \subsetneq C$ for which $D(B) \notin C$ when an absent receiver is hit.
 
%However, it is not clear if Option~2 always leads to the overall smaller $|S|$. More importantly, many results in this paper build on the execution of  Option~1 whenever an absent receiver is hit (regardless of whether there exists some $B \subsetneq C$ for which $D(B) \notin C$) as we require the results to hold for any decoding choice.

% Option~1 is provided as it
% \begin{itemize}
%     \item provides a valid lower bound on the optimal broadcast rate.
%     \item simplifies the proofs of some results where the conditions $B \subsetneq C$ and $D(B) \notin C$ need not be checked.
%     \item can be optimal for certain pliable-index-coding instances, for example, Theorem~\ref{theorem:incomplete} and Corollary~\ref{corollary:3-absent}.
% \end{itemize}

% The above observation leads to the following simplified algorithm, which avoids the need to check for the presence of $B \subsetneq C$ for which $D(B) \notin C$.
% \begin{definition}
%   A simplified version of Algorithm~\ref{algo:2}, or the Simplified Algorithm in short, always chooses Option~1 when it hits an absent receiver, that is, $C \notin \mathbb{U}$.
% \end{definition}

\begin{remark} \label{remark:decoding-chain}
We highlight some properties of Algorithm~\ref{algo:2}:
\begin{enumerate}
\item For a fixed $D$, the only freedom in constructing a chain is in choosing skipped messages in Option~1 and the choice of present receiver $B \subsetneq C$ in Option~2. %So, $(C,S)$ is completely determined by $D$ and the choice of skipped messages.
% \item If an absent receiver~$H$ is hit, then subsequently a message~$a \notin H$ will be skipped, and vice versa.
%   % A message $c_i$ in the chain is skipped if and only if receiver $\{c_1, c_2, \dotsc, c_{i-1}\}$ is absent.
%   So, we skip a message iff we hit an absent receiver.
\item The choice of the skipped message $a \in [1:m] \setminus C$ in line~10 of the algorithm is arbitrary.
\item If a message $a$ is skipped, then the algorithm must have hit an absent receiver, say $H \notin \mathbb{U}$, where $a \notin H$. The converse is not necessarily true because if the algorithm hits an absent receiver, there is a possibility that Option~2 can be executed, where a message need not be skipped. %\PS{this item does not read well.}\LO{
    %re-worded.}
  \item The algorithm always commences by hitting receiver~$\emptyset$ first.
\end{enumerate}
\end{remark}

We note that to obtain a lower bound stated in Theorem~\ref{theorem:chain-lower-bound}, one needs to run Algorithm~\ref{algo:2} over $\prod\limits_{{H \in \mathbb{U}}} \Big|[1:m] \setminus  H \Big|$ possible decoding choices, which is computationally prohibitive for large $m$ and $|\mathbb{U}|$. However, Theorem~\ref{theorem:chain-lower-bound} forms the basis for deriving explicit lower bounds for multiple classes of pliable-index-coding instances in the remainder of this paper, while circumventing the need to evaluate all possible decoding choices.

\section{Other Lower Bounds Based on Theorem~\ref{theorem:chain-lower-bound}}\label{sec:advanced}

In the first two subsections, we derive two lower bounds based on Theorem~\ref{theorem:chain-lower-bound}, while circumventing the need to run Algorithm~\ref{algo:2} over all possible decoding choices. These lower bounds depend only on how the absent receivers form a chain of subsets. In the last subsection, we derive a condition on a collection of absent receivers for which upon hitting any absent receiver therein in Algorithm~\ref{algo:2}, we are guaranteed to be able to avoid skipping a message (i.e., not increasing $|S|$ in Theorem~\ref{theorem:chain-lower-bound}) even if we subsequently hit another absent receiver in the collection. These results will be used subsequently in Section~\ref{sec:application} to derive optimal rates for various scenarios.

\subsection{Nested Chains of Absent Receivers}

\begin{lemma} \label{lemma:necessary-nested}
  If a realisation of Algorithm~\ref{algo:2} skips $L \in \mathbb{Z}^+$ messages,
  then there exists a \textit{nested chain} of absent receivers of at least length $L$, that is, $H_1 \subsetneq H_2 \subsetneq \cdots \subsetneq H_\ell$, for some $\ell \geq L$, with each $H_i \in \absent$. 
\end{lemma} 

\begin{IEEEproof}
  A decoding chain $C$ is constructed by adding messages one by one. So, any receiver that is hit must contain all previously hit receivers. From Remarks~\ref{remark:decoding-chain}, we know that if Algorithm~\ref{algo:2} skips $L$ messages, it must have hit $L$ absent receivers, and these absent receivers must form a nested chain.
\end{IEEEproof}

%The above result leads to the following lower bound, which is easier to use compared to Theorem~\ref{theorem:chain-lower-bound} in some scenarios (for example, case~2 in Theorem~\ref{theorem:nesting} and Theorem~\ref{theorem:perfectly-nested}).
Define $L_\text{max} \in \mathbb{Z}^+$ be the maximum length of any nested chain constructed from receivers absent in a pliable-index-coding instance. Theorem~\ref{theorem:chain-lower-bound} and Lemma~\ref{lemma:necessary-nested} yield the following lower bound, which is a function of only $L_\text{max}$ and the number of messages, $m$, and is independent of the decoding choice~$D$:

\begin{theorem} \label{theorem:simplier-lower-bound}
For any pliable-index-coding instance $\mathcal{P}_{m,\mathbb{U}}$,
\begin{equation}
\beta_q(\mathcal{P}_{m,\mathbb{U}}) \geq m-L_\text{max}.
\label{eq:previous-lower-bound-longest-chain}
\end{equation}
\end{theorem}

\begin{IEEEproof}
  The largest number of skipped messages, evaluated over all decoding choices~$D$ and skipped-message sets, must be upper-bounded by $L_\text{max}$. Otherwise, from Lemma~\ref{lemma:necessary-nested}, we have a nested chain of absent receivers of length $L_\text{max}+1$, which is a contradiction. Thus, 
%  \begin{align*}
$\displaystyle  m- L_\text{max} \leq m - \max_D  \max_{(C,S) \in \mathbb{C}} |S| \leq m - \max_D \min_{(C,S) \in \mathbb{C}} |S| \stackrel{\eqref{eq:chain-lower-bound-new-2}}{\leq} \beta_q(\mathcal{P}_{m,\mathbb{U}}).$
%  \end{align*}   
%  The lower bound in Theorem~\ref{theorem:chain-lower-bound} can be found by evaluating $|S|$ for all decoding choices $D$ and skipped-message sets. Suppose to the contrary that there is a specific decoding choice and a skipped-message set such that $|S| = L+1$. By Proposition~\ref{prop:necessary-nested}, there must be nested chain of absent receivers of length $L+1$, which contradicted the premise of the lemma. So, $\max_D  \min_{(C,S) \in \mathbb{C}} |S| \leq L$. This completes the proof.
\end{IEEEproof}

%Note that the lower bound in Lemma~\ref{theorem:simplier-lower-bound} is actually $\beta_q(\mathcal{P}_{m,\mathbb{U}}) \geq m - \max_D  \max_{(C,S) \in \mathbb{C}} |S|$, which means that it is possibly weaker than that in Theorem~\ref{theorem:chain-lower-bound}.

Theorem~\ref{theorem:simplier-lower-bound} builds on Theorem~\ref{theorem:chain-lower-bound}, but it avoids the need to search over all possible decoding choices~$D$. 
%Therefore, it can be less computationally intensive. %\PS{Can we comment on the computational complexity of finding $L_{max}$?}
%A brute-force method to calculate $L_\text{max}$ has a time complexity of $m \sum_{i=1}^{m/2} \binom{m}{i} = \mathcal{O}(m2^{m})$.

\subsection{Nested Chains with Breakage} \label{sec:new-nested-chain}
 
 From \eqref{eq:chain-lower-bound-new-2}, we see that any upper bound on $L^*$ provides a lower bound on $\beta_q$. For instance, see lower bound~\eqref{eq:previous-lower-bound-longest-chain}, where $L^* \leq L_\text{max}$. 
The lower bound based on $L_\text{max}$ may be loose, as there may exist certain choices of messages skipping that can prevent the algorithm from ever hitting some absent receivers in the longest chain subsequently. Consequently, the algorithm avoids skipping $L_\text{max}$ in total. We now prove another lower bound using this idea. The lower bound is based on the following condition being met:

\begin{condition} \label{condition:1}
($L$-chain breakable) A pliable-index-coding instance is said to be $L$-chain breakable, for some positive integer $L \in [2:m-1]$ if
every chain of absent receivers of length $L$ has the following property. Consider one such chain and denote it by $H_1 \subsetneq \dotsm \subsetneq H_{L}$. There exists some $k \in [1:L-1]$ and some $a \in [1:m] \setminus H_k$ such that there exists no chain of absent receivers of length $L-k$ that contains $H_k \cup \{a\}$, meaning that it is impossible to construct  $\displaystyle (H_k \cup \{a\}) \subseteq H'_{k+1}  \subsetneq \dotsm \subsetneq H'_{L}$, where $H_i' \in \absent$ for all $i \in [k+1:L]$. Here, $H_k \cup \{a\}$ can be either absent or present.
\end{condition}

%\LO{A remark to include multiple break points.}

%\LO{Some trellis algorithm to construct a tree, which determines the largest absent chain from that point.}

The condition above means that we are able to \textit{break} every chain of length $L$, because there exists an appropriate message to skip, such that any continuation from that breakpoint must result in an overall absent-receiver chain of length strictly less than $L$ in total.

\begin{figure}[t]
  \centering
  \includegraphics[scale=0.5]{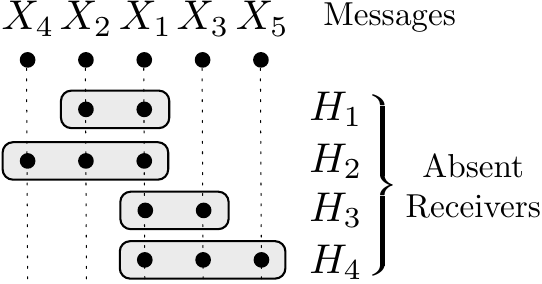}
  \caption{Pliable-index-coding instance $\mathcal{P}_1$ in Examples~\ref{ex:condition-1} and \ref{ex2}}
  \label{fig:ex2}
\end{figure}

\begin{example} \label{ex:condition-1}
Consider a pliable-index-coding instance $\mathcal{P}_1$ with five messages and four absent receivers $H_1=\{1,2\}, H_2=\{1,2,4\}, H_3=\{1,3\}$, and $H_4=\{1,3,5\}$, as depicted in Fig.~\ref{fig:ex2}. $\mathcal{P}_1$ is 2-chain breakable %meets Condition~\ref{condition:1} for $L=2$, 
by considering all chains of absent receivers with length two:
  \begin{itemize}
  \item $H_1 \subsetneq H_2$: Choose $k=1$ and $a = 3$. $H_1 \cup \{3\} = \{1,2,3\}$ is not contained in any absent receiver.
  \item $H_3 \subsetneq H_4$: Choose $k=1$ and $a = 4$. $H_3 \cup \{4\} = \{1,3,4\}$ is not contained in any absent receiver.
  \end{itemize}
\end{example}

Recall that the lower bound given in Theorem~\ref{theorem:simplier-lower-bound} is a function of the longest absent-receiver chain. This lower bound can be tightened as follows if every absent-receiver chain of length $L < L_\text{max} - 1$ can be \textit{broken} in the sense of Condition~\ref{condition:1}.

\begin{theorem} \label{theorem:improved-nested-chain} If $\mathcal{P}_{m,\mathbb{U}}$ is $L$-chain breakable for some positive integer $L \in [2:m-1]$,
%If Condition~\ref{condition:1} holds for some $L \in \mathbb{Z}^+$, 
then $\beta_q (\mathcal{P}_{m,\mathbb{U}}) \geq  m - L + 1$.

%   \begin{quote}
%   Condition: For every chain of absent receivers $H_1\subsetneq H_2 \subsetneq \dotsm \subsetneq H_{L'}$ (for some $L' \geq L$), there exist $H_k \cup \{a\}$ (for some $k \in [1:L-1]$ and for some $a \notin H_k$) that is not contained in any nested chain of $L-k$ absent receivers.
% \end{quote}
% or this one

\end{theorem}
 
\begin{IEEEproof}[Proof of Theorem~\ref{theorem:improved-nested-chain}]
We will show that for any decoding choice, we can always find a realisation of the algorithm that skips at most $L-1$ messages. Then, invoking Theorem~\ref{theorem:chain-lower-bound} completes the proof. 

If every realisation of the algorithm skips at most $L-1$ messages, the result follows trivially.
Otherwise, there exists a realisation, say \textsf{R1}, of the algorithm that skips $L$ or more messages. It follows from Remark~\ref{remark:decoding-chain} that for each skipped message, \textsf{R1} must have hit an absent receiver. Let the sequence of the first $L$ of these absent receivers hit be $H_1 \subsetneq \dotsm \subsetneq H_L$. Since $\mathcal{P}_{m,\mathbb{U}}$ is $L$-chain breakable, there exists some $k \in [1:L-1]$ and $a \in [1:m] \setminus H_k$ such that there exists no chain of absent receivers of length $L-k$ that contains $H_k \cup \{a\}$. 
There always exists another realisation, say \textsf{R2}, which is the same as \textsf{R1} up to the point when it hits the absent receiver $H_k$. At that point, \textsf{R2} has already skipped $k-1$ messages, and it skips message $a$. The $L$-breakable condition dictates that \textsf{R2} will never hit $L-k$ absent receivers subsequently. Therefore, \textsf{R2} skips at most $L-1$ messages. This completes the proof.
\end{IEEEproof}

The following result follows directly from Theorem~\ref{theorem:improved-nested-chain} by searching for the shortest chain
breakability:

\begin{corollary}\label{corrolary:break}
    If $L'$ is the smallest integer for which $\mathcal{P}_{m,\mathbb{U}}$ is $L'$-chain breakable, then $\beta_q (\mathcal{P}_{m,\mathbb{U}}) \geq  m - L' + 1$.
\end{corollary}

We will show the utility of Theorem~\ref{theorem:improved-nested-chain} using the following example:

\begin{example}\label{ex2}
  Consider again the pliable-index-coding instance $\mathcal{P}_1$ in Fig.~\ref{fig:ex2}. The length of the longest nested absent-receiver chain is 2. Theorem~\ref{theorem:simplier-lower-bound} yields $\beta_q \geq 5-2 =3$. 
  From Example~\ref{ex:condition-1}, we know that $\mathcal{P}_1$ is 2-chain breakable.
  %meets Condition~\ref{condition:1} with $L=2$.
Invoking Theorem~\ref{theorem:improved-nested-chain}, we have $\beta_q \geq 5-(2-1)=4$. This lower bound can be achieved by the code~$(X_3+X_5,\, X_1,\, X_2,\, X_4)$ and is therefore tight.
\end{example}

\subsection{Avoid Skipping Messages with Look Ahead} \label{sec:look-ahead}
Continuing with the idea in the previous section, we will show that skipping a particular message can guarantee that we will subsequently not need to skip any message when we hit any member of a special subset of absent receivers. In other words, if any of this special subset of absent receivers is hit subsequently, line~12 (Option~2) of the algorithm can be executed.

\begin{lemma}\label{lemma:look-ahead-skip}
  Suppose there exists $\mathbb{A} \subseteq \absent$ satisfying either of the following conditions:
\begin{enumerate}
\item\label{case1} $\mathbb{A}$ does not cover $[1:m]$, that is, $\mathop{\bigcup}_{H' \in \mathbb{A}} H' \neq [1:m]$.
%\item[]\hspace*{-4ex} For the rest of the cases, $\mathop{\bigcup}\limits_{S' \in \mathbb{S}} S' = [1:m]$.
\item\label{case2} $\mathbb{A}$ is a minimal cover\footnote{A family of sets $\mathbb A = \{A_\ell: \ell \in L\}$ is a minimal cover of $B$ iff (i) $\mathop{\bigcup}_{\ell\in L} A_\ell = B$ and (ii) $\mathop{\bigcup}_{\ell \in L'} A_\ell \subsetneq B$ for any strict subset $L' \subsetneq L$.} of $[1:m]$ and $\bigcap_{H' \in \mathbb{S}} H' \notin \absent$ for some $\mathbb{S} \subseteq \mathbb{A}$, that is, the message intersection of a subset of $\mathbb A$ is present.
\end{enumerate}
  
  Then, if an absent receiver is hit during an instance of Algorithm~\ref{algo:2}, there always exists a message to be skipped such that subsequently as the algorithm proceeds, either (i)~none of the absent receivers in $\mathbb{A}$ will be hit, or (ii)~if any absent receiver in $\mathbb{A}$ is hit, Option~2 is always viable and the algorithm can proceed without skipping any message in that iteration.
\end{lemma}

\begin{IEEEproof}[Proof of Lemma~\ref{lemma:look-ahead-skip}]
Let $C$ be the decoding chain when the algorithm hits the absent receiver. 

  Condition~\ref{case1}:  If $C \subseteq \left( \mathop{\bigcup}_{H' \in \mathbb{A}} H' \right)$, by skipping any $a \in [1:m] \setminus \left( \mathop{\bigcup}_{H' \in \mathbb{A}} H' \right)$, the algorithm will not hit any absent receiver in $\mathbb{A}$. Otherwise, $C$ already contains a message outside $\left( \mathop{\bigcup}_{H' \in \mathbb{A}} H' \right)$, meaning that none of the absent receivers in $\mathbb{A}$ will be hit regardless of what message the algorithm skips now.

  Condition~\ref{case2}: Define $T :=\bigcap_{H' \in \mathbb{S}} H' \notin \absent$ for some $\mathbb{S} \subseteq \mathbb{A}$ such that receiver~$T$ is present, and so $D(T)$ is defined. There must exist an absent receiver~$H_1 \in \mathbb{S}$ that does not contain $D(T)$; otherwise, all absent receivers in $\mathbb{S}$ contain $D(T)$, giving $D(T) \in T$, which is a contradiction. As $\mathbb{A}$ is a minimal cover, there exists some $a \in H_1$ that is not in all other absent receivers in $\mathbb{A}$, that is, $a \notin \mathop{\bigcup}_{H' \in \mathbb{A} \setminus H_1} H'$. 
  \begin{itemize}
      \item If $C \subseteq \mathop{\bigcup}_{H' \in \mathbb{A} \setminus H_1} H'$, then by skipping $a$, the algorithm will never hit any receiver in $\mathbb{A} \setminus H_1$.
      \item Otherwise, $C$ already contains a message outside $\mathop{\bigcup}_{H' \in \mathbb{A} \setminus H_1} H'$, meaning that none of the absent receivers in $\mathbb{A} \setminus H_1$ will be hit regardless of what message the algorithm skips now.
  \end{itemize}
  If the algorithm subsequently hits $H_1$ (that is, $C$ is subsequently updated to $C = H_1$), Option~2 in the algorithm is possible without needing to skip any message, since $T \subsetneq H_1$ and $D(T) \notin C = H_1$.
\end{IEEEproof}

The following example illustrates an application of Lemma~\ref{lemma:look-ahead-skip}:

\begin{example} \label{ex:look-ahead}
    Consider the instance $\mathcal{P}_2$ of the pliable-index-coding instance with six messages. All receivers are present except receivers $H_1=\{3\}$, $H_2=\{1,2,3,4\}$, and $H_3=\{3,4,5,6\}$. 

For this instance, the longest absent receiver chain  $L_\text{max} =2$, which can be obtained from $H_1 \subsetneq H_2$ or $H_1 \subsetneq H_3$. So, Theorem~\ref{theorem:simplier-lower-bound} gives $\beta_q(\mathcal{P}_2) \geq 6-2=4$.

For Theorem~\ref{theorem:improved-nested-chain} to give a better lower bound, Condition~\ref{condition:1} must be satisfied for $L=2$.  However, consider any of the absent-receiver chains of length $L=2$: $H_1 \subsetneq H_2$ or $H_1 \subsetneq H_3$. It is not possible to break both the chains when $k=1$ because no matter what $a \notin H_1$ we choose, we either get $\{a\} \cup H_1 \subsetneq H_2$ or $\{a\} \cup H_1 \subsetneq H_3$.

Furthermore, $\mathcal{P}_2$  does not belong to any instances that have been previously solved~\cite{liutuninetti17}. If we remove all receivers each having at least one and up to four messages, then we get a complement-consecutive complete-$\mathcal{S}$ instance $\mathcal{P}_2^-$ with $\mathcal{S}=\{0,5\}$.
It has been shown~\cite{liutuninetti17} that $\beta_q(\mathcal{P}_2^-) = \min\{m,|S| \} = \min \{6, 2\} = 2$. Alternatively, we can also remove all receivers with zero, one, or five messages to get a complement-consecutive complete-$\mathcal{S}$ instance $\mathcal{P}_2'$ with $\mathcal{S}= [s_\text{min}:s_\text{max}] = [2:3]$. It has been shown~\cite{liutuninetti17} that $\beta_q(\mathcal{P}_2') = \min\{s_\text{max} + 1, m - s_\text{min} \} = 4$.
Invoking Lemma~\ref{lemma:monotonicity}, $\beta_q(\mathcal{P}_2) \geq \max \{\beta_q(\mathcal{P}_2^-), \beta_q(\mathcal{P}_2') \} = 4$.

 We observe that the only way to hit two absent receivers is to first hit $H_1$. When this happens, we invoke Lemma~\ref{lemma:look-ahead-skip} with $\mathbb{A} = \{H_2, H_3\}$, where Condition~\ref{case2} is satisfied with $\mathbb{S} = \mathbb{A}$ and $T = H_2 \cap H_3 = \{3,4\}$.
 Lemma~\ref{lemma:look-ahead-skip} says that there is a message skipping which will guarantee that the algorithm can avoid skipping more messages subsequently (even if it hits $H_2$ or $H_3$).
%$D(\{3,4\})$ can be skipped, such that even if any of $H_2$ or $H_3$ is subsequently hit, Option~2 of the algorithm can be chosen to avoid skipping another message. 
The algorithm will then end with one skipped message, $|S|=1$, regardless of the decoding choices. So, Theorem~\ref{theorem:chain-lower-bound} gives $\beta_q(\mathcal{P}_2) \geq 5$. We will show later in Corollary~\ref{corollary:3-absent} that this bound is tight.

\end{example}

%%%%
%%%%

\section{Applications of Results -- Optimal Rates for Special Cases}\label{sec:application}
%%%%
%%%%

We now derive $\beta_q$ for a few classes of pliable index coding using the lower bounds derived up to this point. These cases are characterised by the absent receivers and are summarised in Table~\ref{table}.

% \usepackage{booktabs}
\begin{table*}[t]
\centering
\caption{Optimal rate results}
\label{table}
{%
\renewcommand{\arraystretch}{0.8}
\begin{tabular}{@{}p{7cm}ll@{}}
\toprule
Scenarios                                                                               & Results                                                                                     & Optimal codes                       \\ \midrule
All absent receivers together do not cover the message\\ ~~set                             & Theorem~\ref{theorem:incomplete}                                                            & Uncoded and cyclic code             \\
No absent receiver                                                                     & Theorem~\ref{thm:no-absent}                                                                 & Uncoded                             \\
No nested absent-receiver pair or just one nested\\ ~~absent-receiver pair                 & Theorem~\ref{theorem:nesting}                                                               & Uncoded and cyclic code             \\
The absent receivers form a special nesting structure\\ ~~called \textit{perfectly nested} & Theorem~\ref{theorem:perfectly-nested}                                                      & Uncoded and cyclic codes            \\
A variation of the perfectly-nested condition\\ ~~called \textit{slightly imperfectly nested}                                      & Theorem~\ref{theorem:less-perfect}                                                          & Uncoded and cyclic codes            \\
A variation of the perfectly-nested condition\\ ~~ called \textit{truncated nested}                                         & Theorem~\ref{theorem:truncated-nested}                                                      & Uncoded, cyclic codes, linear codes \\
Four or fewer absent receivers                                                         & Corollary~\ref{cor:1-and-2-absent}, Theorems~\ref{theorem:3-absent}--\ref{theorem:4-absent} & Uncoded and cyclic code             \\ \bottomrule
\end{tabular}
}
\end{table*}

As shown in Table~\ref{table}, all optimal-rate results in this paper are attained using combinations of uncoded and \textit{cyclic codes} (they are a special case of MDS codes to be defined next), except for Theorem~\ref{theorem:truncated-nested}, which additionally uses specially constructed linear codes. In contrast, existing optimal rate results (that is, consecutive or complement-consecutive complete-$S$ instances) are attained by either uncoded or MDS codes, but not a combination of them.

As an abuse of terminology, we define \textit{cyclic codes} as follows, where the codewords are constructed by cyclic shifts of (the addition of) two messages.
\begin{definition}
    A cyclic code over messages $\{X_1, X_2, \dotsc, X_L\}$ is $(X_1 + X_2, X_2 + X_3, \dotsc, X_{L-1} + X_L) \in \mathbb{F}_q^{L-1}$. For notational convenience, we let the cyclic code over a single message $X_i$ be nil (that is, sending nothing).
\end{definition}

\subsection{Optimal Rates for Extreme Cases}

\begin{theorem} \label{theorem:incomplete}
 Let $\mathcal{P}_{m,\mathbb{U}}$ be such that
  \begin{equation}
   \textstyle \mathop{\bigcup}\limits_{H \in \absent} H \neq [1:m]. \label{eq:union-not-full}
  \end{equation}
  Then $\beta_q(\mathcal{P}_{m,\mathbb{U}}) = m-1$.
\end{theorem}

\begin{IEEEproof}
We invoke Lemma~\ref{lemma:look-ahead-skip} (Condition~\ref{case1}) by setting $\mathbb{A} = \absent$. It follows that Algorithm~\ref{algo:2} will skip at most one message. 
  Invoking Theorem~\ref{theorem:chain-lower-bound}, we have $\beta_q(\mathcal{P}^-) \geq m-1$. This completes the lower bound.

  For achievability, pick any $H \in \absent$. We send $X_H$ uncoded, and $X_{[1:m]\setminus H}$ using a cyclic code. This gives a codelength of $m-1$. Note that any receiver that does not have all messages in $H$ as side information will be able to decode a new message. Also, any receiver that has all messages in $H$ must also have at least one (but not all) messages in $[1:m] \setminus H$---because receiver $H$ is absent---and hence it can decode a new message from the cyclic code.
\end{IEEEproof}

It has been shown~\cite{liutuninetti17} that if all receivers are present, then $\beta_q = m$. The rate of $m$ can be trivially achieved by sending the $m$ messages uncoded. We now strengthen this result to if and only if.
\begin{theorem}\label{thm:no-absent}
  $\beta_q(\mathcal{P}_{m,\mathbb{U}}) = m$ iff $\mathbb{U} = 2^{[1:m]}\setminus \{[1:m]\}$.
\end{theorem}

\begin{IEEEproof}
   We only need to prove the ``only if'' part. Equivalently, we show that if $\mathbb{U} \neq 2^{[1:m]}\setminus \{[1:m]\}$, then $\beta_q(\mathcal{P}_{m,\mathbb{U}}) \neq m$. We start by observing that if $\mathbb{U} \neq 2^{[1:m]}\setminus \{[1:m]\}$, then at least one receiver must be absent. Denoting this absent receiver by $H$, we must have $\mathbb{U} \subseteq 2^{[1:m]}\setminus \{ [1:m], H\} :=\mathbb{U}^+$. For the pliable-index-coding instance $\mathcal{P}_{m,\mathbb{U}^+}$, sending the following code of length $m-1$ satisfies all receivers: $X_H$ uncoded, and $X_{[1:m]\setminus H}$ using a cyclic code. This completes the proof, as
$\beta_q(\mathcal{P}_{m,\mathbb{U}}) \leq \beta_q (\mathcal{P}_{m,\mathbb{U}^+}) \leq m-1$, where the first inequality follows from Lemma~\ref{lemma:monotonicity}, and the second inequality, from the existence of a pliable index code of length $m-1$.
\end{IEEEproof}

%We now present our results to absent receivers $\absent$ in some cases where $\mathop{\bigcup}_{H \in \absent} H = [1:m]$.

\subsection{Optimal Rates for Zero or One Nested Absent-Receiver Pair}

\begin{theorem} \label{theorem:nesting}
  Consider a pliable-index-coding instance $\mathcal{P}_{m,\mathbb{U}}$ where $\mathop{\bigcup}_{H \in \absent} H = [1:m]$. If any of the following is true, then $\beta_q(\mathcal{P}_{m,\mathbb{U}}) = m-1$.
  \begin{enumerate}
  \item (no nested absent pair) $J \nsubseteq K$, for all distinct $J, K \in \absent$.
  \item (one nested absent pair) $J \subsetneq K$, for exactly one pair of $J, K \in \absent$.
  \end{enumerate}
\end{theorem}

\begin{IEEEproof}
%Theorem~\ref{theorem:incomplete} covers the case $\mathop{\bigcup}_{H \in \absent} H \neq [1:m]$. So, in the proof, we consider only $\mathop{\bigcup}_{H \in \absent} H = [1:m]$.
  For achievability, we use the coding scheme for Theorem~\ref{theorem:incomplete}, that is, we choose any $H \in \absent$, and then send $X_H$ uncoded, and $X_{[1:m]\setminus H}$ using a cyclic code. This gives a code of length $m-1$. Note that this code works for the case where only receiver~$H$ is missing, and it will therefore work for the case where $H$ and more receivers are missing.

  For lower bounds, we start with case~1. Since no pair of absent receivers are nested, using Theorem~\ref{theorem:simplier-lower-bound}, we obtain the required lower bound $\beta_q(\mathcal{P}_{m,\mathbb{U}}) \geq m-1$. %This completes the proof for case~1.

For case~2, as there is a pair of nested absent receivers, Theorem~\ref{theorem:simplier-lower-bound} gives a loose lower bound of $m-2$. Suppose that receiver~$\emptyset$ is absent, then $J=\emptyset$, and only one other receiver $K$ can be absent, since the presence of any other absent receiver will yield at least two pairs of nested absent receivers. In this setting then, $\mathop{\bigcup}_{H \in \absent} H = \emptyset \cup K \neq [1:m]$, and by Theorem~\ref{theorem:incomplete}, we see that $\beta_q(\mathcal{P}_{m,\mathbb{U}}) \geq m-1$. 

Now, suppose that case 2 holds and $\emptyset$ is present. With $\emptyset \in \mathbb{U}$, we know that Algorithm~\ref{algo:2} must start without skipping the first message to be included in $C$. We split the decoding choices into three sub-cases, and skip specific messages to avoid $|S|=2$.\\
  \underline{Sub-case 1:} $D$ for which the decoding chain does not hit any absent receiver. For this case, $|S|=0$.\\
  \underline{Sub-case 2:} $D$ for which the decoding chain first hits any absent receiver $H \neq J$. After skipping one message, the algorithm will not hit another absent receiver again, since every receiver that has $H$ as a subset is present (since the only nested pair is $J \subsetneq K$).  This gives $|S|=1$.\\
  \underline{Sub-case 3:} $D$ for which the decoding chain first hits $J$. Then, it is possible to skip $a \in [1:m] \setminus K$. After this, we will not hit another absent receiver, as every receiver that has $J \cup \{a\}$ as a subset is present. This results in $|S|=1$. \\
  Maximising $|S|$ over all $D$, we get the lower bound $\beta_q(\mathcal{P}_{m,\mathbb{U}}) \geq m-1$.
\end{IEEEproof}

\subsection{Optimal Rates for Perfectly $L$-Nested Absent Receivers}

For the next result, we need to first define a class of pliable-index-coding instances.
\begin{definition}
  Given an integer $L \in [1:m-1]$, a pliable-index-coding instance is said to have \textit{perfectly $L$-nested absent receivers} iff the messages $[1:m]$ can be partitioned into $L+1$ subsets $P_0, P_1, \dotsc, P_{L}$ (that is, $\mathop{\bigcup}_{i=0}^L P_i = [1:m]$ and $P_i \cap P_j = \emptyset$ for all $i \neq j$), such that only $P_0$ can be an empty set, and there are exactly $2^L-1$ \textit{absent} receivers, which are
  \begin{equation}
 \textstyle    H_Q := P_0 \cup \left( \mathop{\bigcup}\limits_{i \in Q} P_i \right), \text{ for each } Q \subsetneq [1:L]. \label{eq:q-notation}
  \end{equation} 
\end{definition}

\begin{figure}[t]
  \centering
  \includegraphics[width=0.9\linewidth]{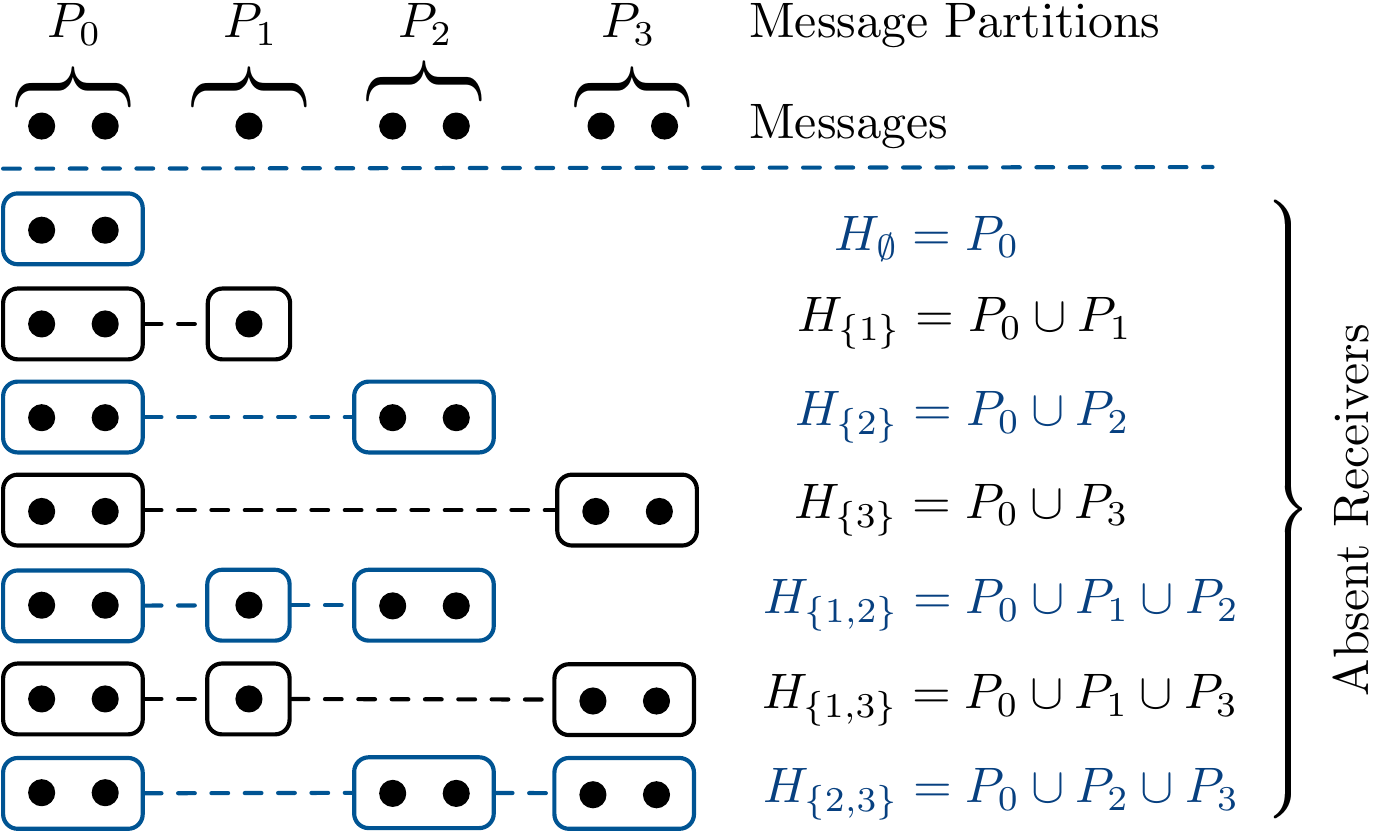}
  \caption{Perfectly 3-nested absent receivers, where there are seven messages (noted by circles), four partitions, and seven absent receivers}
  \label{fig:nested}
\end{figure}

Fig.~\ref{fig:nested} depicts an example of perfectly 3-nested absent receivers.

% Furthermore, for any $Q \subsetneq [1:L]$, define
% \begin{equation}
%  \textstyle    H_Q := P_0 \cup \left( \mathop{\bigcup}\limits_{i \in Q} P_i \right). \label{eq:q-notation}
% \end{equation} 

\begin{theorem} \label{theorem:perfectly-nested}
  For any pliable-index-coding instance $\mathcal{P}_{m,\mathbb{U}}$ with perfectly $L$-nested absent receivers, $\beta_q(\mathcal{P}_{m,\mathbb{U}}) = m-L$.
\end{theorem}

\begin{IEEEproof}
  For achievability we send $X_{P_0}$ uncoded and $X_{P_i}$ for each $i \in [1:L]$ using a cyclic code. One can verify that the decodability of each present receiver is satisfied.

  Since the maximum length of any nested chain of absent receivers is $L$, Theorem~\ref{theorem:simplier-lower-bound} gives  $\beta_q(\mathcal{P}_{m,\mathbb{U}}) \geq m-L$.
\end{IEEEproof}

\subsubsection{Criticality of Perfectly $L$-Nested Absent Receivers}

We will proceed to show that the receivers present in the perfectly $L$-nested absent receiver setting are \textit{critical} in the sense that they are a maximal receiver set that a broadcast rate can support. 

To that end, we introduce the notion of \textit{critical} receivers for pliable index coding. For index coding, it is well-known that removing any message from the side information of any receiver cannot decrease the optimal broadcast rate $\beta$. Hence, the side-information sets of all receivers are said to be critical if removing any messages therein results in a strictly larger $\beta$~\cite{tahmasbishahrasbigohari15}.

However, for pliable index coding, removing messages from side-information sets may increase or decrease $\beta$. We will establish this in Lemma~\ref{lemma:criticality} later. Hence, criticality should not be defined for the messages in side-information sets. However, we can define criticality for pliable index coding with respect to the receivers. For a pliable-index-coding instance $\mathcal{P}_{m,\mathbb{U}}$, the set of receivers $\mathbb{U}$ is said to be \textit{critical} iff adding any receiver to $\mathbb{U}$ strictly increases $\beta_q$.

\begin{lemma}\label{lemma:critical}
If $\absent$ is a set of perfectly $L$-nested absent receivers, then $\mathbb{U}$ is critical.
\end{lemma}

\begin{IEEEproof}
  Start with $\mathcal{P}_{m,\mathbb{U}}$ with perfectly $L$-nested absent receivers.
  We need to show that if we augment $\mathbb{U}$ with any receiver $H = P_0 \cup ( \mathop{\bigcup}_{i \in Q} P_i )$ for some $Q \subsetneq [1:L]$, then $\beta_q(\mathcal{P}_{m,\mathbb{U}^+}) \geq m-L+1$, where $\mathbb{U}^+ = \mathbb{U} \cup H$ is the  set of present receivers after augmenting $\mathbb{U}$ with $H$.

   Imposed by the structure of $\mathbb{U}^+$, for any realisation of Algorithm~\ref{algo:2} to skip $L$ messages, the sequence of absent receivers hit must be of the form 
   \begin{equation}
       H_{Q_0} \subsetneq H_{Q_1} \subsetneq \dotsm \subsetneq H_{Q_{L-1}}, \label{eq:L-chain}
   \end{equation}
   where $|Q_j| = j$ for all $j \in [0:L-1]$, and $Q_{i-1} \subsetneq Q_i$, for all $i \in [1:L-1]$.
  
  Clearly, if the augmented receiver is $H_\emptyset$, then \eqref{eq:L-chain} is not possible. For any decoding choice, any realisation of Algorithm~\ref{algo:2} can only hit at most $L-1$ absent receivers, and hence can skip at most $L-1$ messages. Using Theorem~\ref{theorem:chain-lower-bound}, we have $\beta_q(\mathcal{P}_{m,\mathbb{U}^+}) \geq m-L+1$.

  Otherwise, without loss of generality, let the augmented receiver be $H_{[1:\ell]}$ for some $\ell \in [1:L-1]$. We will show that for any decoding choice, there exists a realisation where the algorithm skips at most $L-1$ messages. To this end, we will show that whenever the algorithm hits an absent receiver, it is possible for the algorithm to skip a specific message such that \eqref{eq:L-chain} is possible only if the sequence of the first $\ell+1$  absent receivers hit is $H_\emptyset \subsetneq H_{[1:1]} \subsetneq H_{[1:2]} \subsetneq \dotsm \subsetneq H_{[1:\ell]}$. Since $H_{[1:\ell]}$ is present, we get the desired result.

  More specifically, consider the following realisations. Algorithm~\ref{algo:2} starts by hitting $H_\emptyset$, it skips a message in $P_1$. For \eqref{eq:L-chain} to hold, we only need to consider decoding choices that lead to the algorithm next hitting the absent receiver $H_{[1:1]}$.\footnote{All other decoding choices will not result in \eqref{eq:L-chain}, and hence the algorithm will skip at most $L-1$ messages.} After hitting $H_{[1:1]}$, the algorithm skips a message in $P_2$. Again, for \eqref{eq:L-chain} to hold, we only need to consider decoding choices that lead to the algorithm next hitting $H_{[1:2]}$. We repeat the same argument (that is, whenever the algorithm hits $H_{[1:i]}$, it skips a message in $P_{i+1}$) until the algorithm hits $H_{[1:\ell-1]}$. It then skips a message in $P_\ell$. Since $H_{[1:\ell]}$ is present (being augmented in $\mathbb{U}^+$), only one of the two following groups of realisations can happen for any decoding choice:
  \begin{itemize}
      \item The next absent receiver being hit is $H_{Q}$ for some $|Q| \in [\ell+1:L-1]$.
      \item The algorithm does not hit any more absent receiver.
  \end{itemize}
  In any group, \eqref{eq:L-chain} is not possible.

  So, for every decoding choice, we can always find a realisation (with specific skipped messages) where the total number of skipped messages is at most $L-1$. Theorem~\ref{theorem:chain-lower-bound} gives $\beta_q(\mathcal{P}_{m,\mathbb{U}^+}) \geq m-L+1$.
\end{IEEEproof}

\subsubsection{Reducing Side Information May Increase or Decrease the Optimal Broadcast Rate}
With the above results, we show the following characteristic of pliable index coding, which is in stark contrast to index coding:
\begin{lemma}\label{lemma:criticality} 
For any pliable-index-coding instance $\mathcal{P}_{m,\mathbb{U}}$, removing a message from a present receiver $H \in \mathbb{U}$ may strictly increase or strictly decrease the optimal broadcast rate $\beta_q(\mathcal{P}_{m,\mathbb{U}})$.
\end{lemma}

We prove Lemma~\ref{lemma:criticality} using the example below:

\begin{example} \label{example:increase-decrease}
  Consider $m=5$ and a set of absent receivers~$\absent_1 = \{ \{1,2,3\}, \{3\}, \{3,4\}\}$. Using Theorem~\ref{theorem:incomplete}, we have $\beta_q(\mathcal{P}_{m,\mathbb{U}_1}) = m-1$. Now, we remove message~5 from a present receiver~$\{3,4,5\} \in \mathbb{U}_1$. This is equivalent to replacing the present receiver~$\{3,4,5\}$ with a new present receiver~$\{3,4\}$. We get $\absent_2 = \{ \{1,2,3\}, \{3\}, \{3,4,5\}\}$, which forms perfectly 2-nested absent receivers. Using Theorem~\ref{theorem:perfectly-nested}, $\beta_q(\mathcal{P}_{m,\mathbb{U}_2}) = m-2$. We continue by removing messages 2 and 4 from the present receiver $\{2,3,4\} \in \mathbb{U}_2$.  This replaces the present receiver $\{2,3,4\}$ with a new present receiver $\{3\}$. We get $\absent_3 = \{ \{1,2,3\}, \{2,3,4\}, \{3,4,5\}\}$. Using Theorem~\ref{theorem:nesting}, $\beta_q(\mathcal{P}_{m,\mathbb{U}_3}) = m-1$.
\end{example}

 \subsection{Optimal Rates for the Slightly Imperfectly $L$-Nested Absent Receivers}

% We have previously defined a class of pliable-index-coding problems as follows~\cite{ongvellambikliewer2019}:
% \begin{definition}
%   A pliable-index-coding problem is said to have \textit{perfect $L$-nested absent receivers} iff the messages $[1:m]$ can be partitioned into $L+1 \in [2:m]$ subsets $P_0, P_1, \dotsc, P_{L}$ (that is, $\mathop{\bigcup}_{i=0}^L P_i = [1:m]$ and $P_i \cap P_j = \emptyset$ for all $i \neq j$), such that only $P_0$ can be an empty set, and there are exactly $2^L-1$ \textit{absent} receivers, which are defined as
%   \begin{equation}
%  \textstyle    H_Q := P_0 \cup \left( \mathop{\bigcup}_{i \in Q} P_i \right), \text{ for each } Q \subsetneq [1:L].
%   \end{equation}
% \end{definition}

% % \begin{figure}[t]
% %   \centering
% %   \includegraphics[scale=0.9]{43}
% %   \caption{Perfect $3$-nested absent receivers}
% %   \label{fig:perfect}
% % \end{figure}
% % Fig.~\ref{fig:perfect} depicts an example of perfect $3$-nested absent receivers.
% For any pliable-index-coding problem $\mathcal{P}_{m,\mathbb{U}}$ with perfect $L$-nested absent receivers, $\beta_q(\mathcal{P}_{m,\mathbb{U}}) = m-L$~\cite{ongvellambikliewer2019}.

We now prove the optimal rate for pliable-index-coding instances with slightly imperfectly $L$-nested absent receivers, defined as follows:
\begin{definition}
    Given an integer $L \in [1:m-1]$, a pliable-index-coding instance $\mathcal{P}_{m, \mathbb{U}_1}$ is said to have \textit{slightly imperfectly $L$-nested absent receivers} iff it differs from a pliable-index-coding instance $\mathcal{P}_{m, \mathbb{U}_2}$ with perfectly $L$-nested absent receivers as follows: There exist one $H$ and one $\tilde{H} \subsetneq H$ such that
    \begin{itemize}
        \item $H \notin \mathbb{U}_2$ and $H \in \mathbb{U}_1$, and
        \item $\tilde{H} \in \mathbb{U}_2$ and $\tilde{H} \notin \mathbb{U}_1$.
    \end{itemize}
    In other words, we make a receiver~$H$ that is absent in $\mathbb{U}_2$ to be present in $\mathbb{U}_1$, and also make a receiver~$\tilde{H} \subsetneq H$ that is present in $\mathbb{U}_2$ to be absent in $\mathbb{U}_1$. Doing this does not change the total number of present (or absent) receivers, that is, $|\mathbb{U}_1| = |\mathbb{U}_2|$.
    \end{definition}

\begin{figure}[t]
  \centering
    \includegraphics[width=0.9\linewidth]{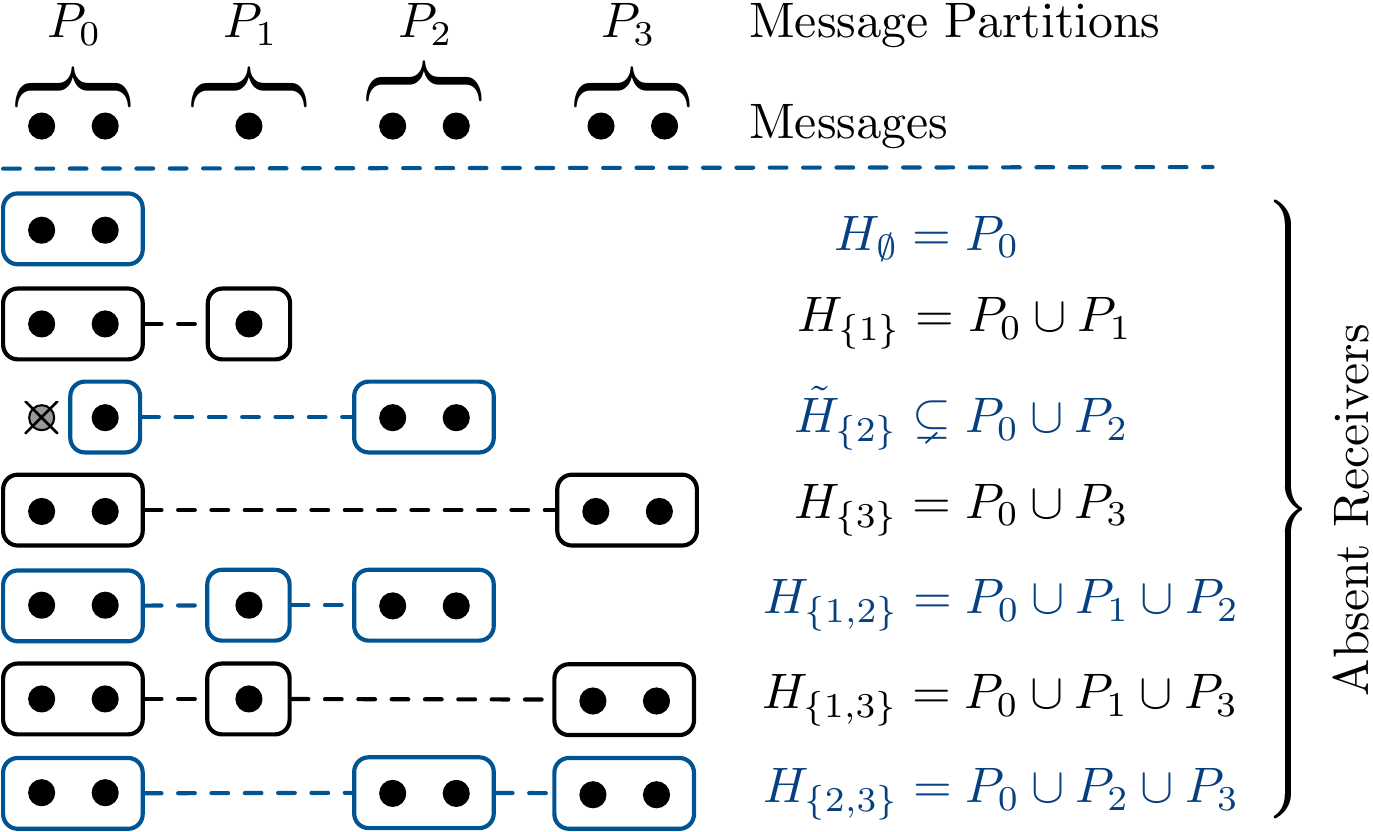}
  \caption{Slightly imperfectly $3$-nested absent receivers, formed by replacing one absent receiver with a present receiver of a smaller side-information set in the perfectly $3$-nested setup depicted in Figure~\ref{fig:nested}}
  \label{fig:imperfect}
\end{figure}

Fig.~\ref{fig:imperfect} depicts an example of slightly imperfectly $3$-nested absent receivers.

\begin{theorem}\label{theorem:less-perfect}
    For any pliable-index-coding instance $\mathcal{P}_{m,\mathbb{U}}$ with slightly imperfectly $L$-nested absent receivers, $\beta_q(\mathcal{P}_{m,\mathbb{U}})  = m - L + 1$.
\end{theorem}

\indent\indent \textit{Proof:}  See the Appendix.

Theorem~\ref{theorem:less-perfect} gives the following corollary:
\begin{corollary}\label{corollary:3-absent}
        Consider a pliable-index-coding instance $\mathcal{P}_{m, \mathbb{U}}$, where the set of absent receivers is $\absent = \{H_1, H_2, H_3\}$, such that $H_1 \subsetneq H_2 \cap H_3$, and $H_2 \cup H_3 = [1:m]$. We have $\beta_q(\mathcal{P}_{m, \mathbb{U}}) = m-1$.
\end{corollary}
\begin{IEEEproof}
  $\mathcal{P}_{m,\mathbb{U}}$ is formed by having perfectly 2-nested absent receivers with $P_0 = H_2 \cap H_3$, $P_1 = H_2 \setminus H_3$, $P_2 = H_3 \setminus H_2$, and then replacing absent receiver $H_{\emptyset} = P_0$ with $\tilde{H}_{\emptyset} = H_1 \subsetneq P_0$. Using Theorem~\ref{theorem:less-perfect}, we have $\beta_q(\mathcal{P}_{m,\mathbb{U}})  = m - 1$.
%  Let $S= H_1$ and $\mathbb{S} = \{ H_2, H_3\}$. Invoking Lemma~\ref{lemma:look-ahead-skip} (Case~\ref{case2}), the maximum number of skipped messages $L^* \leq 1$, and substituting it into \eqref{eq:lower-bound-L-star} gives $\beta_q(\mathcal{P}_{m,\mathbb{U}}) \geq m-1$. The lower bound can be achieved by sending $X_{H_2}$ uncoded and the rest using a cyclic code.
\end{IEEEproof}

The optimal rate for Example~\ref{ex:look-ahead} can be directly obtained by invoking Corollary~\ref{corollary:3-absent}.

\subsection{Optimal Rates for $T$-Truncated $L$-Nested Absent Receivers}

We define another variation of perfectly $L$-nested absent receivers.
\begin{definition} \label{def:truncated}
  Given two integers $L \in [1:m-1]$ and $T \in [0:L-1]$, a pliable-index-coding instance is said to have \textit{$T$-truncated $L$-nested absent receivers} iff the messages $[1:m]$ can be partitioned into $L+1$ subsets $P_0, P_1, \dotsc, P_{L}$ (that is, $\mathop{\bigcup}_{i=0}^L P_i = [1:m]$ and $P_i \cap P_j = \emptyset$ for all $i \neq j$), such that only $P_0$ can be an empty set, and there are $\sum_{i=0}^T \binom{L}{i}$ absent receivers, which are defined as 
  \begin{equation}
 \textstyle    H_Q = P_0 \cup \left( \mathop{\bigcup}_{i \in Q} P_i \right),\;\; \forall Q \subsetneq [1:L],\text{ with } |Q| \in [0:T].
\end{equation}
\end{definition}

\begin{figure}[t]
  \centering
  \includegraphics[width=0.9\linewidth]{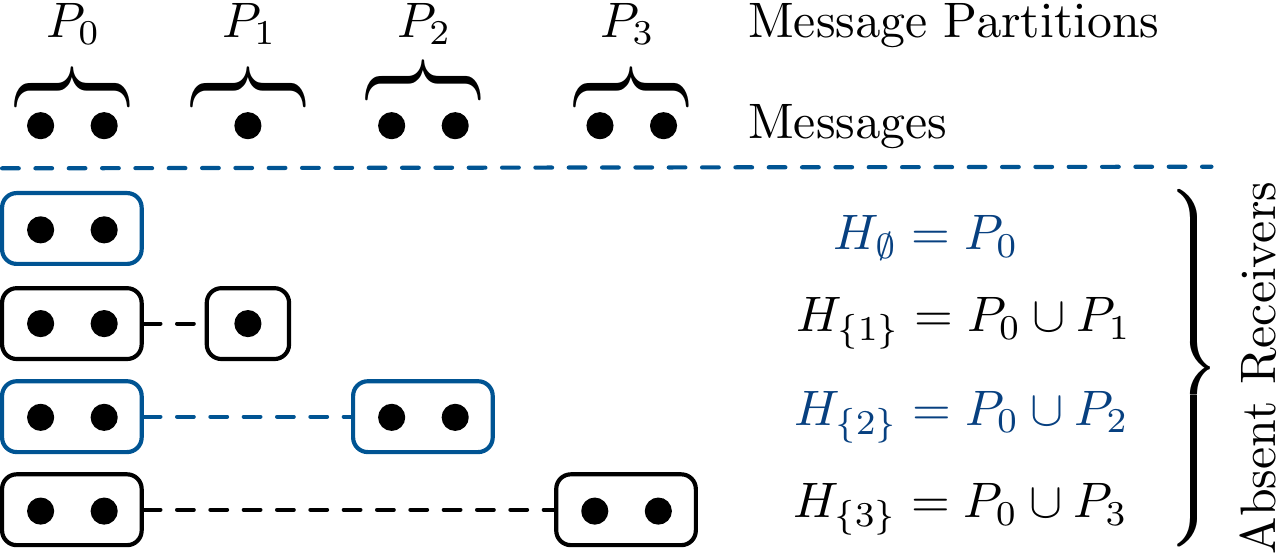}
  \caption{$1$-truncated $3$-nested absent receivers, formed by retaining only the upper four absent receivers in the perfectly $3$-nested setup depicted in Figure~\ref{fig:nested} 
  }
  \label{fig:truncated}
\end{figure}

Note that $(L-1)$-truncated $L$-nested absent receivers are equivalent to perfectly $L$-nested absent receivers. Fig.~\ref{fig:truncated} depicts an example of $1$-truncated $3$-nested absent receivers.

\begin{theorem} \label{theorem:truncated-nested}
  For any pliable-index-coding instance $\mathcal{P}$ with $T$-truncated $L$-nested absent receivers, $\beta(\mathcal{P}) = \beta_q(\mathcal{P}) = m-T-1$, for sufficiently large $q$ in general and for any $q$ if $T \in \{L-2, L-1\}$.
\end{theorem}

\indent\indent \textit{Proof:}  See the Appendix.

\subsection{Optimal Rates for Up To Four Absent Receivers}

From Theorem~\ref{thm:no-absent}, we have established that  $\beta_q = m$ if and only if there is no absent receiver, that is $|\absent|=0$.

Using Theorems~\ref{theorem:incomplete} and \ref{theorem:nesting}, we further establish the following:
\begin{corollary} \label{cor:1-and-2-absent}
  If $1 \leq |\absent| \leq 2$, then $\beta_q=m-1$.
\end{corollary}
\begin{IEEEproof}
  For $|\absent|=1$, by definition, the absent receiver $H \subsetneq [1:m]$, and hence $\mathop{\bigcup}_{H \in \absent} H \neq [1:m]$. The result follows from Theorem~\ref{theorem:incomplete}.
For $|\absent|=2$,
if $\mathop{\bigcup}_{H \in \absent} H \neq [1:m]$, we again have $\beta_q=m-1$ from Theorem~\ref{theorem:incomplete}. Otherwise, $\mathop{\bigcup}_{H \in \absent} H = [1:m]$, and since
there can be either no nested pair or one nested pair of absent receivers, the result follows from Theorem~\ref{theorem:nesting}.
\end{IEEEproof}

Next, we proceed to characterise the optimal rate with three and four absent receivers.

\begin{theorem} \label{theorem:3-absent}
  Suppose $|\absent|=3$. Then
  \begin{equation*}
    \beta_q =
    \begin{cases}
      m-2, & \text{ if the absent receivers are perfectly 2-nested},\\
      m-1, & \text{ otherwise}.
    \end{cases}
  \end{equation*}
\end{theorem}

\begin{theorem}\label{theorem:4-absent}
  Suppose $|\absent|=4$. Then
  \begin{equation*}
    \beta_q =
    \begin{cases}
      m-2, & \text{ if a subset of absent receivers is either}\\ & \text{ perfectly 2-nested or 1-truncated 3-nested},\\
      m-1, & \text{ otherwise}.
    \end{cases}
  \end{equation*}
\end{theorem}

%\LO{A table listing different cases for each $m = 3, 4, 5$}

\indent\indent \textit{Proofs of Theorems~\ref{theorem:3-absent} and \ref{theorem:4-absent}:}  See the Appendix.

\section{Summary}

% Recaps the main contributions,
% Emphasizes any surprising or particularly elegant results,
% Mentions possible directions for future work.

We proposed a novel method for deriving lower bounds on the optimal rate of pliable index coding. Unlike most approaches in communication theory that focus on the receivers present in the system, our technique characterises lower bounds based on the properties of absent receivers. For absent receivers whose side-information sets satisfy certain structural or nesting conditions, we have shown that the resulting lower bounds are tight. Future work could explore broader classes of side-information structures for which the lower bounds remain tight or can be further improved.

%For the cases where $|\absent| \leq 4$, we see that only structured case results in savings beyond one coded symbol.

%\section{Conclusions}

\bibliography{mypublications,otherpublications}

\appendix

\begin{IEEEproof}[Proof of Theorem~\ref{theorem:less-perfect}]
  We will use Algorithm~\ref{algo:2} and show that $L^*  \leq L-1$. Note that the length of any longest chain of nested absent receivers in $\mathcal{P}_{m,\mathbb{U}}$ is $L$. %, and any of these chain must be in the form $H_\emptyset \subsetneq H_{A_1} \subsetneq H_{A_2} \subsetneq \dotsc \subsetneq H_{A_{L-1}}$, for some distinct $\emptyset \subsetneq A_1 \subsetneq \dotsm \subsetneq A_{L-1} \subsetneq [1:L]$, where $|A_i|=i$.
  
  We will now consider all decoding choices $D$ that can potentially result in $L$ absent receivers being hit. We need not consider any other decoding choices, as we will hit at most $L-1$ absent receivers and skip at most $L-1$ messages.

    Let $\tilde{H}_Q \subsetneq P_0 \cup \left( \mathop{\bigcup}_{i \in Q} P_i \right)$ be the imperfect absent receiver. 
  If the imperfect absent receiver is $\tilde{H}_\emptyset \subsetneq P_0$, then we only need to prove the case where the first receiver to be hit is $\tilde{H}_\emptyset$; otherwise, we will hit at most $L-1$ absent receivers. We invoke Lemma~\ref{lemma:look-ahead-skip} (Condition~\ref{case2}), where
  $\mathbb{S} = \mathbb{A} =  \Big\{ (P_0  \cup  P_\ell): \ell \in [1:L] \Big\}$, and  $T = P_0 \notin \absent$. To hit $L$ absent receivers in total after hitting $\tilde{H}_\emptyset$, the next absent receiver to be hit must be from $\mathbb{A}$. From Lemma~\ref{lemma:look-ahead-skip}, we know that we need not skip any message when we hit any absent receiver in $\mathbb{A}$. So, in total, we will skip at most $L-1$ messages.

  Otherwise, consider the imperfect absent receiver being $\tilde{H}_{Q_\ell} \subsetneq H_{Q_\ell}$, where $|Q_\ell|=\ell$ for some $\ell \in [1:L-1]$. Without loss of generality, let $Q_\ell = [1:\ell]$.
 When the algorithm hits $H_{Q_i}$, where $|Q_i|=i$ for $i \in [0:\ell-2]$, it is always possible for the algorithm to skip a message in $P_{i+1}$. In such realisations, for the algorithm to ever skip $L$ messages, the first $\ell$ absent receivers hit must be in this order: $H_\emptyset \subsetneq H_{\{1\}} \subsetneq H_{[1:2]} \subsetneq \dotsm \subsetneq H_{[1:\ell-1]}$.

Note that the imperfect absent receiver is $\tilde{H}_{[1:\ell]} \subsetneq \mathop{\bigcup}_{i \in [0:\ell]} P_i $. Among $\{P_i: i \in [0:\ell]\}$, we say that $P_j$ is an affected set iff $\tilde{H}_{Q_\ell} \cap P_j \neq P_j$.
When the algorithm hits $H_{[1:\ell-1]}$, the algorithm can skip any message as follows:
    \begin{equation}
      a \in
      \begin{cases}
        P_\ell \setminus \tilde{H}_{Q_\ell}, &\text{if } \tilde{H}_{Q_\ell} \cap P_\ell \neq P_\ell \text{  }\\
        &\text{(that is, $P_\ell$ is one of the affected sets)},\\
        P_\ell, & \text{otherwise.}
      \end{cases} \label{eq:skip-option}
    \end{equation} 
We observe the following:
  \begin{enumerate}
  \item If $\tilde{H}_{Q_\ell} \cap P_\ell \neq P_\ell$ (that is, $P_\ell$ is one of the affected sets), then the algorithm can skip $a \in P_\ell \setminus \tilde{H}_{Q_\ell}$, and then it will not hit $\tilde{H}_{Q_\ell}$.
  \item Otherwise, $\tilde{H}_{Q_\ell} \cap P_j \neq P_j$, for some $j \in [0:\ell-1]$ (that is, $P_j$ is one of the affected sets), the algorithm can skip any $a \in P_\ell$. Since the decoding chain already contains  $(P_0 \cup \dotsm \cup P_{\ell-1})$, the algorithm will also not hit $\tilde{H}_{Q_\ell}$.
  \end{enumerate}
Consequently, the next receiver to be hit can only be $H_{Q_i}$, for some $|Q_i| \geq \ell+1$. So, the total number of absent receivers hit is at most $L-1$.

 We have shown that regardless of which decoding choice, there exists a realisation of the algorithm that skips at most $L-1$ messages, and $L^* \leq L-1$. Hence, $\beta_q(\mathcal{P}_{m,\mathbb{U}})  \geq m - L + 1$.

We now prove achievability.
According to Theorem~\ref{theorem:perfectly-nested}, sending a code of length $m-L$ satisfies all receivers in the perfectly $L$-nested instance. In comparison, this slightly imperfectly nested instance contains a present receiver, $H_Q = P_0 \cup \left( \mathop{\bigcup}\limits_{i \in Q} P_i \right)$, that is not present in the perfectly nested instance. To satisfy this receiver, we transmit another message $X_a$ for some $a \in P_k$ for some $k \in [1:L] \setminus Q$. The length for this code is thus $m-L+1$.  
\end{IEEEproof}

{\ }

\begin{IEEEproof}[Proof of Theorem~\ref{theorem:truncated-nested}]
  It is easy to see that the longest nested chain in this case is $T+1$, and the chain consists of absent receivers $H_{Q_0} \subsetneq H_{Q_1} \subsetneq \dotsm \subsetneq H_{Q_T}$, where $Q_0 \subsetneq Q_1 \subsetneq \dotsm \subsetneq Q_T$ and $|Q_i|=i$ for all $i \in [0:T]$. So, \eqref{eq:previous-lower-bound-longest-chain} gives a lower bound $\beta_q(\mathcal{P}) \geq m-T-1$.
  
  Let the instance with  perfectly $L$-nested absent receivers on the same partitions $\{P_i: i \in [0:L]\}$ be $\mathcal{P}_{m,\mathbb{U}^-}$, where $\mathbb{U}^-$ are the receivers present in the instance.  From Theorem~\ref{theorem:perfectly-nested}, we know that $\beta_q(\mathcal{P}_{m,\mathbb{U}^-}) = m-L$. Achievability for $\mathcal{P}_{m,\mathbb{U}^-}$ is attained by sending messages in $P_0$ uncoded, $X_{P_0} = Y_0 \in \mathbb{F}_q^{|P_0|}$, and messages in each $P_i$, $i \in [1:L]$, using a cyclic code $Y_i = (Z_{i,1}+Z_{i,2},\, Z_{i,2}+Z_{i,3},\, \dotsc,\, Z_{i,|P_i|-1}+Z_{i,|P_i|}) \in \mathbb{F}_q^{|P_i|-1}$, where $Z_{i,j}$ is the $j$th message in $P_i$. Let this code be $Y=(Y_0, \dotsc, Y_L) \in \mathbb{F}_q^{m-L}$.

  We will now add receivers group by group until we get the $T$-truncated $L$-nested instance, with receivers $\mathbb{U}$. At each stage, we compose additional coded messages to satisfy newly added receivers.
  \begin{itemize}
  \item First, we add receivers $H_Q$ with $|Q|=L-1$ to $\mathbb{U}^-$; equivalently, we remove these receivers from the absent receiver set. To satisfy these added receivers (the other original receivers can decode with $Y$ and their side information), we add a coded message $V_{L-1} = \sum_{i=1}^L Z_{i,1} \in \mathbb{F}_q$. Each added receiver knows all but one message in $\{Z_{i,1}: i \in [1:L]\}$, and can then decode a new message from $V_{L-1}$.
  \item Then, we further add receivers $H_Q$ with $|Q|=L-2$. To satisfy the newly added receivers, we add another a coded message $V_{L-2} = \sum_{i=1}^L \gamma^{i} Z_{i,1} \in \mathbb{F}_q$, where $\gamma$ is a primitive element in $\mathbb{F}_q$. Note each newly added receiver knows all but two messages in $\{Z_{i,1}: i \in [1:L]\}$ and can then decode a new message from $(V_{L-1},V_{L-2}) \in \mathbb{F}_q^2$.
  \item This step is repeated. That means when we add receivers $H_Q$ with $|Q| = L- k$, for $k \in [1:L-1-T]$, we also add a coded message $V_{L-k} = \sum_{i=1}^L (\gamma^{k-1})^{i} Z_{i,1} \in \mathbb{F}_q$. Each newly added receiver knows $L-k$ messages in $\{Z_{i,1}: i \in [1:L]\}$, and can then decode a new message from $(V_{L-1}, V_{L-2}, \dotsc, V_{L-k}) \in \mathbb{F}_q^k$ if $q$ is sufficiently large.
  \end{itemize}
So, by sending $(Y_0, Y_1, \dotsc, Y_L, V_{L-1}, V_{L-2}, \dotsc, V_{L-(L-1-T)}) \in \mathbb{F}_q^{m-T-1}$, every receiver in $\mathcal{P}_{m,\mathbb{U}}$ can obtain at least one new message. So, the rate of $m-T-1$ is achievable for sufficiently large $q$. Also, note that if $L-1-T = 1$, then only one additional message $V_{L-1}$ needs to be sent, and this message can be of any field size.
\end{IEEEproof}

{\ }

\begin{IEEEproof}[Proof of Theorem~\ref{theorem:3-absent}]
  Let the absent receivers be $H_1$, $H_2$, and $H_3$, where the labelling is arbitrary. If $\mathop{\bigcup}\limits_{i=1}^3 H_i \neq [1:m]$, then $\beta_q = m-1$ according to Theorem~\ref{theorem:incomplete}.

  For the remaining settings, we have $\mathop{\bigcup}\limits_{i=1}^3 H_i = [1:m]$. For this case, the length of the longest nested chain of absent receivers is at most two. Therefore, there can be at most two pairs of nested absent receivers.
  \begin{itemize}
  \item If there is one or no nested pair of absent receivers, we have
    $\beta_q = m-1$ according to Theorem~\ref{theorem:nesting}.
  \item Otherwise,
    we have two nested absent receiver pairs, and they must have the
    configuration $H_1 \subseteq (H_2 \cap H_3)$, and
    $H_2 \cup H_3 = [1:m]$. The two nested pairs are $H_1 \subsetneq H_2$ and $H_1 \subsetneq H_3$. For this case, we have two scenarios:
    \begin{itemize}
    \item If $H_1 \subsetneq (H_2 \cap H_3)$, then $\beta_q =
    m-1$ according to Corollary~\ref{corollary:3-absent}. 
  \item Otherwise, $H_1 = H_2 \cap H_3$, which is perfectly 2-nested,
    then  $\beta_q = m-2$ according to Theorem~\ref{theorem:perfectly-nested}.
  \end{itemize}

\end{itemize}
The proof is complete by noting that the last case is the only case with perfectly 2-nested absent receivers.
\end{IEEEproof}

{\ }

\begin{IEEEproof}[Proof of Theorem~\ref{theorem:4-absent}]
  Let the absent receivers be $\{H_i: i \in [1:4]\}$, where the labelling is arbitrary. If $\mathop{\bigcup}\limits_{i=1}^4 H_i \neq [1:m]$, then $\beta_q = m-1$ according to Theorem~\ref{theorem:incomplete}.
For the rest of the proof, we only have to focus on the case that $\mathop{\bigcup}\limits_{i=1}^4 H_i = [1:m]$. Note that in this case, since each $H_i\subsetneq [1:m]$, the length of the longest nested chain of absent receivers is $L_\text{max} \leq 3$. We will proceed by considering various cases based on the \emph{minimum cover number}, which is defined to be the size of a minimal cover of absent receivers as defined in Condition~\ref{case2} of Lemma~\ref{lemma:look-ahead-skip}. For each value of the minimum cover number, several subcases based on the overlap among the four absent receivers are identified and analysed. A pictorial representation of each of the cases described below can be found in Figures~\ref{fig:Thm11CasesAB}, \ref{fig:Thm11CaseC}, and \ref{fig:case}.

\begin{figure*}[t]
  \centering
  \includegraphics[scale=0.6]{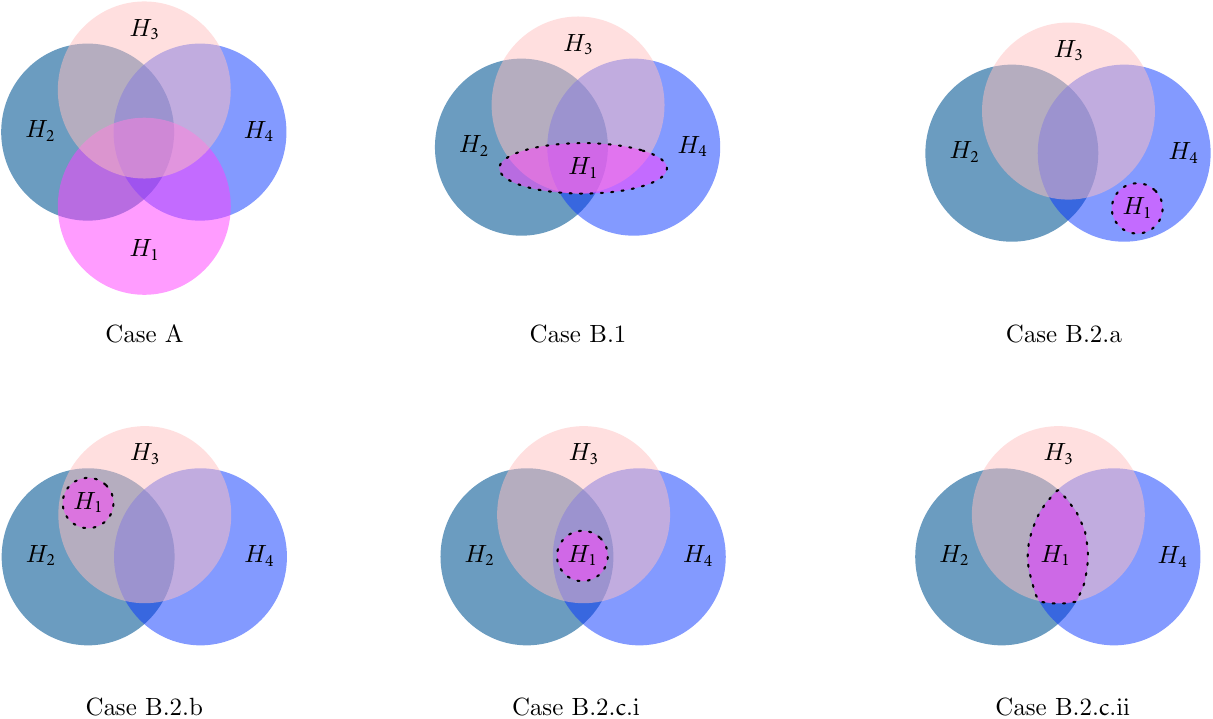}
  \caption{Various nested structures of absent receivers in Cases A and B when the minimum cover number is at least three 
  }
  \label{fig:Thm11CasesAB}
\end{figure*}

\color{black}

% Suppose that $L_\text{max} = 3$. Then, we must have the configuration $H_1 \subsetneq H_2 \subsetneq H_3$ and $H_3 \cup H_4 = [1:m]$, meaning that $(H_3,H_4)$ is not a nested pair. Now, we invoke Theorem~\ref{theorem:improved-nested-chain}, and consider $L=3$ and the chain $H_1 \subsetneq H_2 \subsetneq H_3$. For $k=1$ note that $H_1 \cup \{a\}$, for some $a \notin H_2$, can only be contained in $H_3$ and $H_4$, which are not nested. So, $H_1 \cup \{a\}$ is not contained in any nested chain of $L-k=2$ absent receivers. Theorem~\ref{theorem:improved-nested-chain} gives $L^* \leq L-1=2$. If $L_\text{max} \leq 2$, $L^* \leq 2$ follows immediately.  So, $\beta_q \geq m-2$ for all cases.

\medskip
\noindent  \underline{Case A: The minimum cover number is four}. Then, there is no nested pair of absent receivers, and hence $\beta_q = m-1$ according to Theorem~\ref{theorem:nesting}.

\medskip
  \noindent \underline{Case B: The minimum cover number of $[1:m]$ is three}. WLOG, let $H_2 \cup H_3 \cup H_4 = [1:m]$. Then, $L_\text{max} \leq 2$ where any nested pair, if it exists, will have $H_1$ nested in one or more receivers in $\{H_2, H_3, H_4\}$. The following subcases then arise.
  \begin{enumerate}
  \item If $L_\text{max} =1$, then Theorem~\ref{theorem:simplier-lower-bound} gives
    $\beta_q \geq m-1$. This lower bound is achievable by sending $X_{H_1}$ uncoded and $X_{[1:m]\setminus H_1}$ using a cyclic code. 
  \item Otherwise,
    $L_\text{max}=2$, and so $\beta_q \geq m-2$.
     \begin{enumerate}
    \item If there is only one
    nested pair, then
    $\beta_q = m-1$ according to Theorem~\ref{theorem:nesting}. 
  \item If there
    are two nested pairs, say, WLOG, $H_1 \subsetneq H_2$ and $H_1 \subsetneq H_3$, meaning that $H_1 \subseteq H_2 \cap H_3$ and $H_1 \nsubseteq H_4$.
    The only way to hit two absent receivers is to first hit $H_1$, and
    then hit either $H_2$ or
    $H_3$. When $H_1$ is hit, we can skip any message in $[1:m] \setminus (H_2 \cup H_3) = H_4 \setminus (H_2 \cup H_3)$. By doing this, $|S| \leq 1$, and hence $\beta_q \geq m - 1$. The lower bound is achievable by sending $X_{H_1}$ uncoded and $X_{[1:m]\setminus H_1}$ using a cyclic code.
  \item Otherwise, there are three nested pairs, and we must have $H_1 \subseteq H_2 \cap H_3 \cap H_4$, where the nested pairs are $H_1 \subsetneq H_i$, for $i \in [2:4]$.
    \begin{enumerate}
    \item If $H_1 \subsetneq H_2 \cap H_3 \cap H_4$, we invoke Lemma~\ref{lemma:look-ahead-skip} (Condition~\ref{case2}) by setting $\mathbb{A} = \mathbb{S} =  \{H_2, H_3, H_4\}$ and noting that $H_2 \cap H_3 \cap H_4$ is present. We know that even if Algorithm~\ref{algo:2} hits $H_1$, it is possible to skip a message such that it will never hit any of $H_2$, $H_3$, or $H_4$. So, the maximum number of skipped messages is one. With this, we can again show that $\beta_q \geq m-1$, which is achievable by sending $X_{H_1}$ uncoded and $X_{[1:m]\setminus H_1}$ using a cyclic code. 
    \item Otherwise, $H_1= H_2 \cap H_3 \cap H_4$.
      \begin{enumerate}
      \item[--] If $H_1 = H_i \cap H_j$ for all distinct pairs $i,j \in [2:4]$,  then the family of absent receivers is 1-truncated 3-nested  with $T=1$ and $L=3$. This scenario corresponds to  the one in Fig.~\ref{fig:truncated} by setting $P_0 = H_1$, $P_1 = H_2 \setminus H_1$, $P_2 = H_3 \setminus H_1$, and $P_3 = H_4 \setminus H_1$. Using Theorem~\ref{theorem:truncated-nested}, we get $\beta_q =m-2$. Since $T=L-2$, this result is valid for any field size $q$.
      \item[--] Otherwise, $H_1 \subsetneq H_i \cap H_j$ for some distinct pair $i,j \in [2:4]$. Invoking Lemma~\ref{lemma:look-ahead-skip} (condition~\ref{case2}) with $\mathbb{A} = \{H_2, H_3, H_4\}$ and $\mathbb{S} = \{H_i, H_j\}$, we can again show the lower bound $\beta_q \geq m-1$, which is also achievable by sending $X_{H_1}$ uncoded and $X_{[1:m]\setminus H_1}$ using a cyclic code. 
      \end{enumerate}
 
  \end{enumerate}

  \end{enumerate}

\end{enumerate}

\begin{figure*}[t]
  \centering
  \includegraphics[scale=0.6]{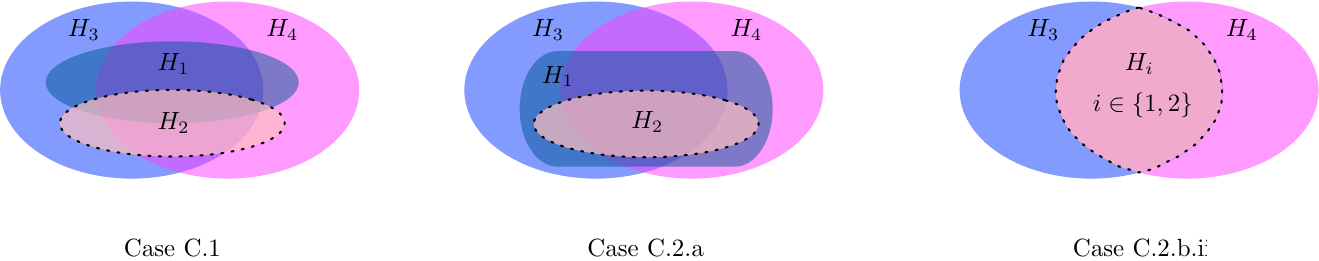}
  \caption{Some subcases of nested structures of absent receivers in Case C when the minimum cover number is two
  }
  \label{fig:Thm11CaseC}
\end{figure*}

\medskip
\noindent \underline{Case C: The minimum cover number of $[1:m]$ is two.} WLOG, let $H_3, H_4$ be two minimal subsets of absent receivers that cover $[1:m]$, i.e., $H_3 \cup H_4 = [1:m]$. The following subcases then arise.
\begin{enumerate}
  \item If $L_\text{max} =1$, then similar to the argument above, $\beta_q =m-1$.
  \item Otherwise $L_\text{max} \in \{2,3\}$, and $\beta_q \geq m-L_\text{max}$.
    \begin{enumerate}
    \item If $H_i \nsubset H_j$ for any $i \in \{1,2\}$ and any $j \in \{3,4\}$, then $\{H_1,H_2\}$ must form the required and the only nested pair. Then
    $\beta_q = m-1$ according to Theorem~\ref{theorem:nesting}.
    \item Otherwise, at least one nested pair involves $H_3$ or $H_4$. For any nested pair that involves $H_3$ or $H_4$, let $H_3$ or $H_4$ be the larger set in the pair. Otherwise, we re-label the sets without affecting the minimum cover, which is $\{ H_3, H_4\}$. %WLOG, let $H_1 \subsetneq H_3$ be the nested pair.
          \begin{enumerate}
            \item \label{case:perfect-2-nested} 
             If $H_i = H_3 \cap H_4$, for any $i \in \{1,2\}$:\\ Then $(H_i, H_3, H_4)$ is perfectly 2-nested. Denote the instance with three absent receivers $\{H_i, H_3, H_4\}$  by $\mathcal{P}^+$, and the original instance with four absent receivers $\{H_1, H_2, H_3, H_4\}$ by $\mathcal{P}$. This gives $\beta_q(\mathcal{P}) \overset{(a)}\leq \beta_q(\mathcal{P}^+) \overset{(b)}= m-2$, where (a) is due to Lemma~\ref{lemma:monotonicity} and (b) is due to Theorem~\ref{theorem:perfectly-nested}.
             
            \medskip
              
              Note that to get a lower bound of $\beta_q \geq m-3$, Algorithm~\ref{algo:2} must hit three absent receivers, with the first one being $H_j$, $j \in \{1,2\}$, and the second one being $H_{\bar{j}}$, where $\bar{j} = ((2 - j) \mod 2 )+ 1$, and $H_j \subsetneq H_{\bar{j}}$. But it is possible to avoid this upon hitting $H_j$, as it is possible to skip a message $a \notin H_{\bar{j}}$. Doing so, the algorithm can only hit $H_3$ or $H_4$ next. Since $H_3 \cup H_4 = [1:m]$, it will not hit another absent receiver anymore. So, $\beta_q \geq m-2$.
           Combining with $\beta_q(\mathcal{P}) \leq m-2$, we get $\beta_q(\mathcal{P}) = m-2$. 
            
            \item Otherwise, $H_i \neq H_3 \cap H_4$, for any $i \in \{1,2\}$, meaning that $H_3 \cup H_4$ is present. If $H_i$ does not cover $H_a \setminus H_b$, for any $i \in \{1,2\}$ and any $a,b \in  \{3,4\}$:\\ In other words, $(H_a \setminus H_b) \nsubseteq H_i$ for every combination of $i \in \{1,2\}$ and $a,b \in \{3,4\}$. WLOG, let the decoding choice of $H_3 \cap H_4$ be $D(H_3 \cap H_4) \in H_b \setminus H_a$. The only way to skip two or more messages is to first hit say $H_i, i \in \{1,2\}$. Using a similar argument in the proof of Lemma~\ref{lemma:look-ahead-skip} (Condition~\ref{case2}) with $\mathbb{A} = \mathbb{S} = \{H_3, H_4\}$, it is possible to skip any message in $H_a \setminus (H_b \cup H_1 \cup H_2)$, for $a,b \in \{3,4\}$. We can always find such a message because $H_i$, for any $i \in \{1,2\}$, does not cover $H_a \setminus H_b$, and if $(H_1 \cup H_2)$ covers $H_a \setminus H_b$, then $(H_1,H_2,H_b)$ would be a minimum cover, which has been dealt with for the cases with a minimum cover of three. Doing so, only $H_a$ can be hit next, and if that happens, the algorithm can avoid skipping a message with decoding choice $D(H_3 \cup H_4) \in H_b \setminus H_a$. This gives $\beta_q(\mathcal{P}) \geq m-1$. This lower bound can be attained by sending $X_{H_1}$ uncoded and $X_{[1:m]\setminus H_1}$ using a cyclic code. 
            
\begin{figure}[t]
  \centering
  \includegraphics[scale=0.6]{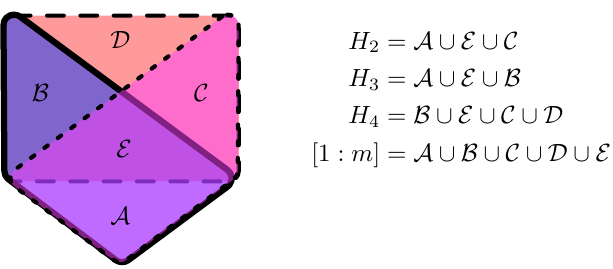}
  \caption{The Venn diagram for $\{H_2, H_3, H_4\}$ for Case C.2.b.iii, where $H_3 \setminus H_4 = \mathcal{A} \subseteq H_2 = \mathcal{A} \cup \mathcal{E} \cup \mathcal{C}$}
  \label{fig:case}
\end{figure}

            \item\label{case} Otherwise, $H_i \neq H_3 \cap H_4$, for any $i \in \{1,2\}$, and $H_i$ covers $H_a \setminus H_b$, for some $i \in \{1,2\}$ and $a,b \in  \{3,4\}$. WLOG, let $H_2$ cover $H_3 \setminus H_4$, or in other words, $H_3 \setminus H_4 \subseteq H_2$:\\ Also, for any nested pair that involves $H_1$, $H_1$ is the smaller set (This is due to the aforementioned premise that $H_3$ or $H_4$ is the larger set in the nesting. If $H_2 \subsetneq H_1$, we swap the labels of 1 and 2.). Fig.~\ref{fig:case} shows the Venn diagram for $H_2$, $H_3$, and $H_4$, where $H_3 \setminus H_4 = \mathcal{A} \neq \emptyset$ and $H_4 \setminus H_3 = \mathcal{C} \cup \mathcal{D} \neq \emptyset$ because $H_3$ and $H_4$ are the minimum cover. Also, $\mathcal{B} \cup \mathcal{C} \neq \emptyset$ because $H_2 \neq H_3$. WLOG, let $\mathcal{B} \neq \emptyset$ (otherwise, we swap the labels of 2 and 3). 

            \medskip
            
            First, we consider $\mathcal{C} \neq \emptyset$ and $\mathcal{D} \neq \emptyset$. For Algorithm~\ref{algo:2} to hit two or more absent receivers, it must first hit $H_1$ (the smaller set), and $H_1$ cannot intersect with both $\mathcal{A}$ and $\mathcal{D}$ (otherwise, there is no nested pair). We now show that there exist realisations of the algorithm that do not skip two messages. Suppose that $H_1$ is hit.\\ 
            -- If $H_1 \cap \mathcal{A} \neq \emptyset$, the algorithm can skip any message in $\mathcal{D}$ and will not hit any absent receiver again.\\ 
            -- Else if $H_1 \cap \mathcal{D} \neq \emptyset$, the algorithm can skip any message in $\mathcal{A}$ and will not hit any absent receiver again.\\ 
            -- Otherwise, $H_1 \cap (\mathcal{A} \cup \mathcal{D}) = \emptyset \Rightarrow H_1 \subseteq (\mathcal{B} \cup \mathcal{C} \cup \mathcal{E})$. We only need to consider the case where receivers $(\mathcal{E} \cup \mathcal{B})$ and $(\mathcal{E} \cup \mathcal{C})$ are present (Otherwise, if $H_1 = \mathcal{E} \cup \mathcal{B} = H_3 \cap H_4$, we obtain case~C.2.b.i; else if $H_1 = \mathcal{E} \cup \mathcal{C} = H_2 \cap H_4$, we swap the label of $H_2$ and $H_3$ and, again, get case~C.2.b.i.). If any of the receivers $(\mathcal{E} \cup \mathcal{B})$ and $(\mathcal{E} \cup \mathcal{C})$ decodes some $x \in \mathcal{A}$, the algorithm can skip any message in $\mathcal{D}$. The only absent receiver that we will next hit is $H_4$, and if that happens, the algorithm can avoid skipping a message. Otherwise,  with decoding choices $D(\mathcal{E} \cup \mathcal{B}) \in (\mathcal{C} \cup \mathcal{D})$ and  $D(\mathcal{E} \cup \mathcal{C}) \in (\mathcal{B} \cup \mathcal{D})$, the algorithm can skip any message in $\mathcal{A}$. The next absent receiver that we can hit is either $H_2$ or $H_3$, and we can always avoid skipping a message when that happens. \\
            Note that the above arguments to avoid skipping still hold even if the message to be skipped is already in the decoding chain (in that case, we can skip any other message).

            \medskip
            
            Now consider $\mathcal{C} = \emptyset$ (which means $\mathcal{D} \neq \emptyset$). If $H_1$ is hit first, we follow the arguments above for $\mathcal{C} \neq \emptyset$ and $\mathcal{D} \neq \emptyset$. The arguments are valid even if $\mathcal{C} = \emptyset$, as long as $\mathcal{A} \neq \emptyset$, $\mathcal{B} \neq \emptyset$, and $\mathcal{D} \neq \emptyset$. Otherwise, $H_2$ is hit first, the algorithm can skip a message in $\mathcal{D} \setminus H_1$ to avoid skipping again (unless $H_1 = \mathcal{A} \cup \mathcal{E} \cup \mathcal{D}$, which is not possible as we assumed that $H_2 \nsubseteq H_1$).

            \medskip

            Lastly, consider $\mathcal{C} \neq \emptyset$ and $\mathcal{D} = \emptyset$.\\
            -- If $H_1 \neq \mathcal{E}$, then we can use Lemma~\ref{lemma:look-ahead-skip} (Condition~\ref{case2}) with $\mathbb{A} = \mathbb{S} = \{H_2, H_3, H_4\}$ to show that if $H_1$ is hit, we can skip a message such that we do not need to skip any more messages. If any of $\{H_2,H_3,H_4\}$ is hit first, we will not hit another receiver subsequently.\\
            -- Otherwise, if $H_1 = \mathcal{E}$, $(H_1, \dotsc, H_4)$ is perfectly 2-nested, which gives $\beta_q = m-2$.

            \medskip

            In summary, except for the perfectly nested case, we can always skip a message such that the total number of skipped messages is at most one, giving $\beta_q \geq m-1$. This lower bound can be achieved by sending $X_{H_1}$ uncoded and $X_{[1:m]\setminus H_1}$ using a cyclic code. 
            \end{enumerate}
            \end{enumerate}
  \end{enumerate}
  \end{IEEEproof}
  \begin{IEEEbiographynophoto}{Lawrence Ong} (Senior Member, IEEE)
  received the Bachelor degree in electrical engineering from the National University of Singapore in 2001, the MPhil degree from the University of Cambridge in 2004, and the PhD degree from the National University of Singapore in 2008. He is currently an associate professor in the School of Engineering at the University of Newcastle, Australia. His research interests include information theory, secure communications, data privacy, and index coding. He was awarded a Discovery Early Career Researcher Award and a Future Fellowship from the Australian Research Council in 2012 and 2014, respectively. He served as an Editor for the IEEE Transactions on Communications from 2018 to 2023.
  \end{IEEEbiographynophoto}

  \begin{IEEEbiographynophoto}{Badri N.\ Vellambi}(Senior Member, IEEE) received the B.Tech. degree in electrical engineering from the Indian Institute of Technology-Madras, Chennai, India, in 2002, and the M.S. degree in electrical engineering, the M.S. degree in mathematics, and the Ph.D. degree in electrical engineering from the Georgia Institute of Technology, Atlanta, GA, USA, in 2005, 2008, and 2008, respectively. He held Post-Doctoral Research Fellowship positions with the Institute for Telecommunications Research, University of South Australia, the New Jersey Institute of Technology, Newark, NJ, USA, and the Research School of Computer Science, Australian National University, Canberra, Australia, from August 2008 to August 2018. He is currently an Associate Professor (Educator) with the Department of Electrical \& Computer Engineering, University of Cincinnati, Cincinnati, OH, USA. His current research interests include information theory, data compression, channel coding, statistical learning, reinforcement learning, and artificial intelligence.
  
  \end{IEEEbiographynophoto}

  \begin{IEEEbiographynophoto}{Parastoo Sadeghi}
  received the bachelor’s and master’s degrees in electrical engineering from the Sharif University of Technology, Tehran, Iran, in 1995 and 1997, respectively, and the Ph.D. degree in electrical engineering from the University of New South Wales (UNSW), Sydney, in 2006. She is currently a Professor with the School of Engineering and Technology, UNSW Canberra. She has co-authored over 220 refereed journal articles and conference papers. Her research interests include information theory, data privacy, index coding, and network coding. From 2016 to 2019 and since April 2025, she has served as an Associate Editor for the IEEE Transactions on Information Theory. In 2022, she was selected as a Distinguished Lecturer of the IEEE Information Theory Society. She is currently serving as a member of the Board of Governors of the IEEE Information Theory Society.
  \end{IEEEbiographynophoto}

  \begin{IEEEbiographynophoto}{J\"{o}rg Kliewer} (Fellow, IEEE) received the Dr. (Ing.) (Ph.D.) degree in electrical engineering from the University of Kiel, Germany, in 1999. From 2000 to 2003, he was a Senior Researcher and a Lecturer at the University of Kiel. In 2004, he visited the University of Southampton, U.K., for one year. From 2005 to 2007, he was with the University of Notre Dame, IN, USA, as a Visiting Assistant Professor. From 2007 to 2013, he was with New Mexico State University, Las Cruces, NM, USA, most recently as an Associate Professor. He is currently with New Jersey Institute of Technology, Newark, NJ, USA, as a Professor, where he is directing the Elisha Yegal Bar-Ness Center of Machine Intelligence, Communication, and Signal Processing. His research interests span information and coding theory, machine learning, generative models, and secure and private communication. He was a recipient of the Leverhulme Trust Award, the German Research Foundation Fellowship Award, the IEEE GLOBECOM Best Paper Award, and a Fulbright Scholarship. He was an Associate Editor and an Area Editor of IEEE Transactions on Communications from 2008 to 2014 and from 2015 to 2021, respectively. He was an Associate Editor of IEEE Transactions on Information Theory from 2017 to 2020. From 2021 to 2023, he served as an Editor for IEEE Transactions on Information Forensics and Security.  
  \end{IEEEbiographynophoto}
  
\end{document}